\documentclass{llncs}
\usepackage{graphicx,floatflt}
\usepackage[lined,boxed,commentsnumbered,ruled,vlined,noend,linesnumbered,boxed]{algorithm2e}
\usepackage{subfig}
\newcommand{\remove}[1]{}
\newtheorem{observation}{Observation}
\usepackage{color}

\newcommand{\IV}{\textsf{\sc Inplace\_Visibility}}
\newcommand{\RV}{\textsf{\sc Readonly\_Visibility}}

\newcommand{\LNV}{\textsf{\sc LeftNonVisible}}
\newcommand{\RNV}{\textsf{\sc RightNonVisible}}
\newcommand{\MLP}{\textsf{\sc Minimum\_Link\_Path}}
\title{Space-efficient Algorithms for Visibility Problems in Simple
Polygon \vspace{-0.1in}}

\author{Minati De\inst{1} \and Anil Maheshwari\inst{2} \and Subhas C. Nandy\inst{1}} 
\institute{Indian Statistical Institute, Kolkata 700108, India \and 
School of Computer Science, Carleton University, Ottawa, Canada}
\begin{document}

\maketitle

\begin{abstract} \vspace{-0.1in}
Given a simple polygon $P$ consisting of $n$ vertices, we study the problem of 
designing space-efficient algorithms for computing (i) the visibility polygon 
of a point inside $P$, (ii) the weak visibility polygon of a line segment inside 
$P$ and (iii) the minimum link path between a pair of points inside $P$. For
problem (i) two algorithms are proposed. The first one is  an in-place algorithm
where the input array may be lost. It uses only $O(1)$ extra space apart from
the input array. The second one assumes that the input is given in a read-only
array, and it needs $O(\sqrt{n})$ extra space. The time complexity of both the
algorithms are $O(n)$. For problem (ii), we have assumed that the input polygon
is given in a read-only array. Our proposed algorithm runs in $O(n^2)$ time
using $O(1)$ extra space. For problem (iii) the time and space complexities of
our proposed algorithm are $O(kn)$ and $O(1)$ respectively; $k$ is the length
(number of links)  in a minimum link path between the given pair of points.
\vspace{-0.2in}
\end{abstract}

\vspace{-0.2in}
\section{Introduction}\vspace{-0.1in}
Visibility is one of the foundational areas in computational geometry and
it has applications in various domains, including robot motion
planning, guarding art galleries, computer graphics, GIS, sensor network. For an
illustrated survey,  see \cite{G07}. Recently,  visibility algorithms are
being embedded in the hardware of digital cameras, sensors, etc, and the constraint
on the size of the instrument has become important. So, the
algorithm designers are now becoming interested in developing space-efficient
algorithms for various  visibility problems.  

So far, in-place algorithms have been  studied for a very few problems in
computational geometry (see \cite{ChanChen2010}). In \cite{AsanoMRW11}, constant
work space algorithms for the following visibility related problems are studied:
(i) triangulation of a simple polygon,
(ii) triangulation of a point set, (iii) Euclidean shortest path between a
pair of points inside a simple polygon, and (iv) Euclidean minimum spanning
tree, where the input is given in a read-only array. The time complexity of the
algorithms for the problems (i)-(iii) are
$O(n^2)$, and that for problem (iv) is $O(n^3)$. The open question was 
whether one can compute the visibility of a point
inside a simple polygon in sub-quadratic time, where the polygon is given in a
read-only array \cite{AsanoMRW11}. Recently, two algorithms for this problem are
proposed by Barba
et al. \cite{Barba}. The first one is deterministic, and it requires 
$O(n\overline{r})$ time and $O(1)$ space, where $\overline{r}$ is the number of
reflex vertices of the output visibility polygon. The second one is a randomized
algorithm and it requires  $O(n\log r)$ time and $O(\log r)$ space, where $r$ is
the number of reflex vertices in the input polygon. 

\noindent {\bf New Results:} In this paper we present the following results:
%We started working in this direction and obtained the following results:
\vspace{-0.1in}
\begin{itemize}
\item[$\bullet$] An in-place algorithm for computing the visibility polygon of a
point inside a simple polygon $P$ in $O(n)$ time and $O(1)$ extra work-space,
where $n$ is the number of vertices of $P$. Note that, after the execution of
the algorithm, the polygon can not be retrieved from the content of the array.
\item[$\bullet$] An $O(\sqrt{n})$ space algorithm for computing the visibility
polygon of a point inside the polygon $P$ in $O(n)$ time, where the vertices of
$P$ are given in a read-only array. 
\item[$\bullet$] An $O(n^2)$ time and $O(1)$ extra work-space algorithm for
computing the weak-visibility of the polygon $P$ from a line segment inside the
polygon, where the vertices of $P$ are given in a read-only array.
\item[$\bullet$] An $O(kn)$ time and $O(1)$ extra work-space algorithm for
computing a minimum link path between a pair of points $s$ and $t$ inside $P$,
where the vertices of $P$ are given in a read-only array and $k$ is the size of
the output.
\end{itemize}

\vspace{-0.25in}
\section{Visibility of a point inside simple polygon}
\vspace{-0.2in}\label{point-vis}
Let the vertices of the polygon $P=\{p_1,p_2,\ldots,p_n\}$ be given in an array
$P$ in anticlockwise order. Initially, $P[i]$ contains the coordinates of the
$i$-th vertex $p_i$ of $P$ in the given order. Let $\pi$ be a given point inside
$P$. We will consider the problem of computing the visibility polygon of $\pi$
in $P$. It is easy to see that the visibility polygon of $\pi$ can be computed
in $O(n^2)$ time and $O(1)$ extra space where the vertices of $P$ are given in a
read-only array. We will describe two algorithms, namely \IV\ and \RV. The first
one computes the visibility polygon in  in-place manner with $O(1)$ extra space.
The second one assumes that the input array is read-only, and it uses
$O(\sqrt{n})$ extra space. The time complexity of both the algorithms are
$O(n)$. 

\vspace{-0.2in}
\subsection{\IV} \label{IV} 
\begin{figure}[h]
\vspace{-0.2in}
\centering
\includegraphics[scale=0.2]{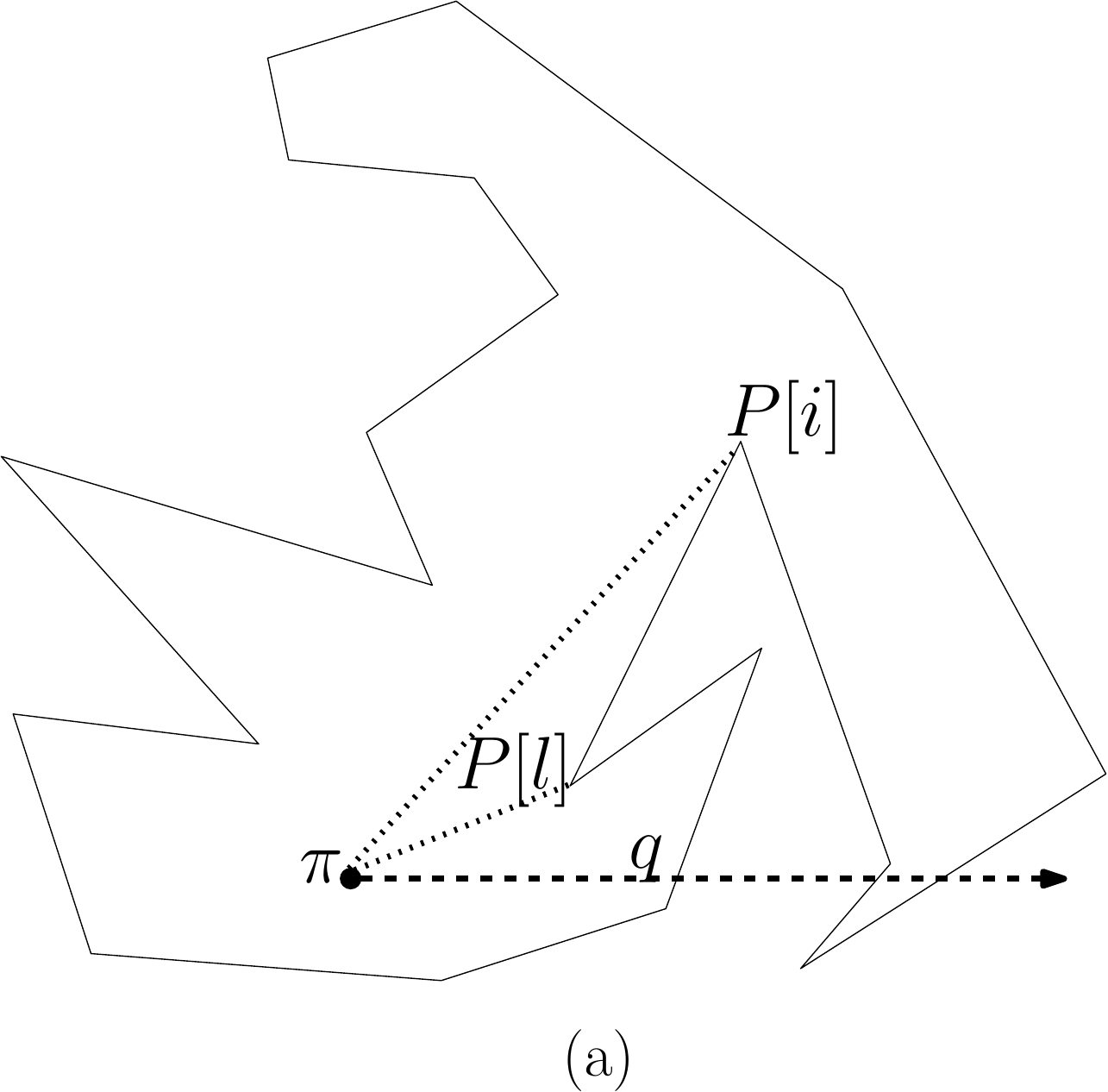}\hspace{0.1in}
\includegraphics[scale=0.2]{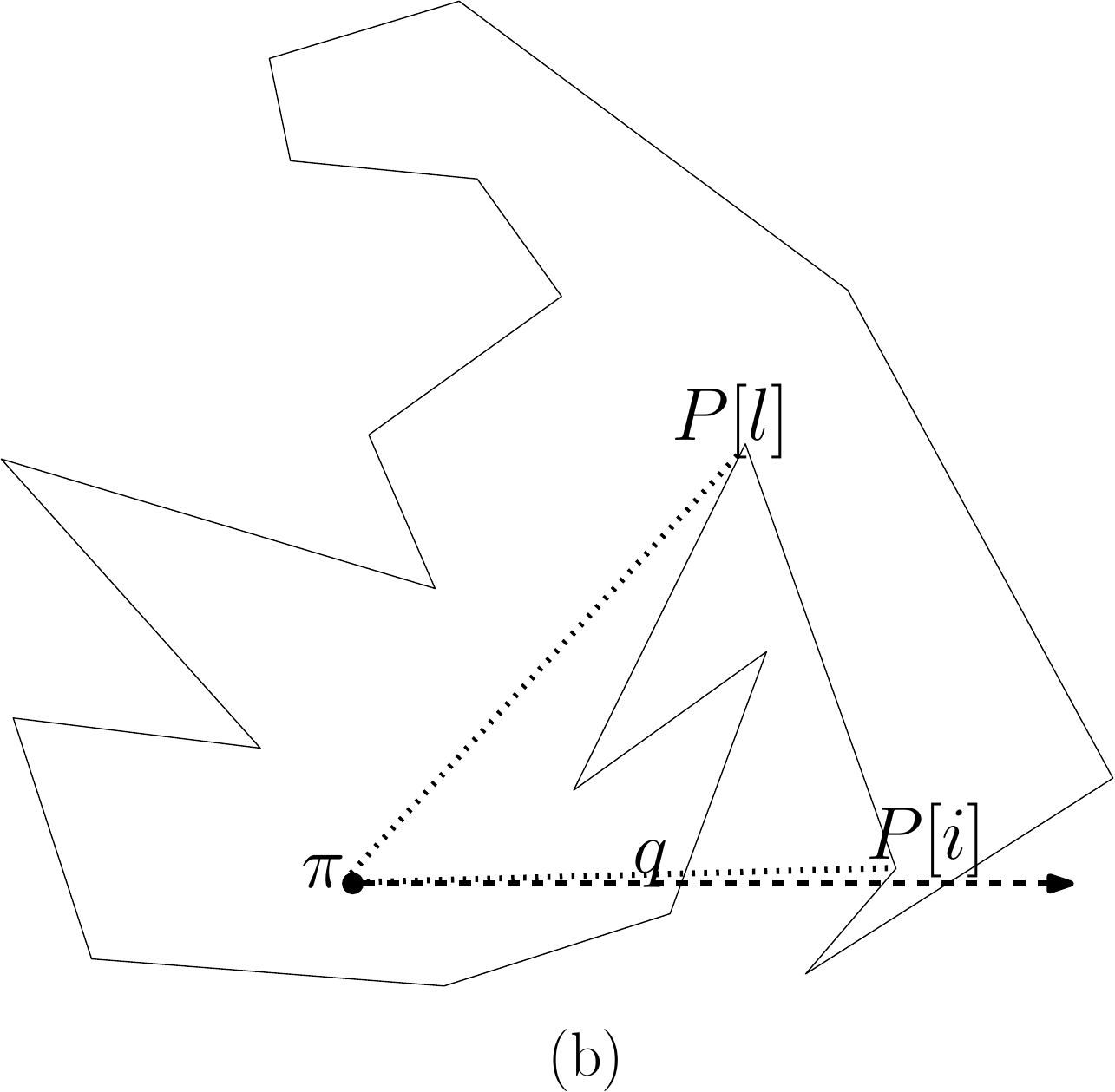}\hspace{0.1in}
\includegraphics[scale=0.2]{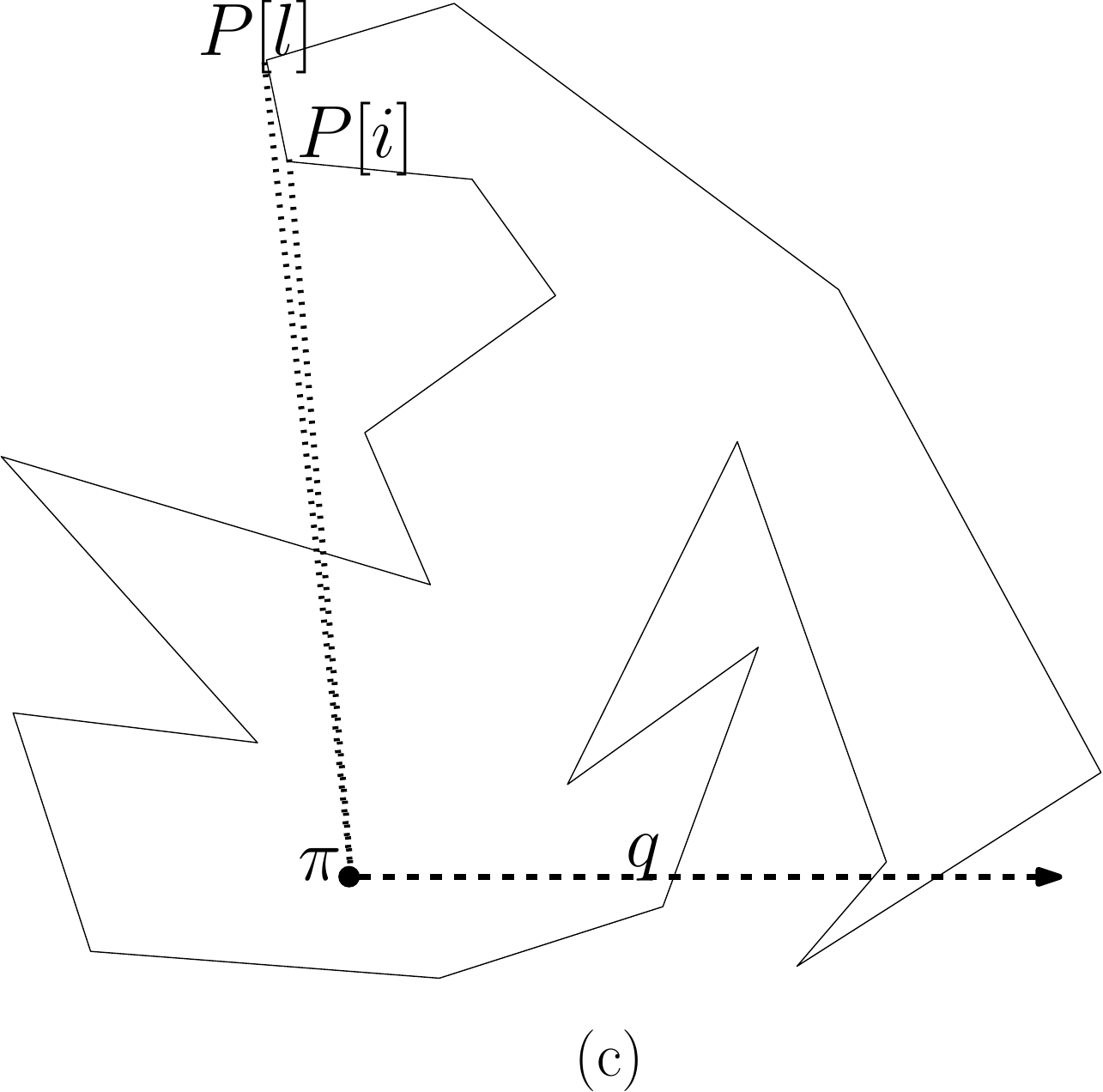}
\caption{ Processing of vertex $P[i]$ }\vspace{-0.3in} 
\label{IVcases}
\end{figure}

\vspace{-0.1in}
We first describe an algorithm to report all the vertices of $P$ that are
visible from $\pi$. Later, we show that the above algorithm can be easily
modified to report the visibility polygon of $\pi$. We start the algorithm by
drawing a horizontal ray  $\overrightarrow{H}$ through the point $\pi$ to its
right that finds an edge $e_\theta=(p_\theta, p_{\theta+1})$ of $P$ intersected
by $\overrightarrow{H}$ first. Let $q$ be the point of intersection
$\overrightarrow{H}$ and $e_\theta$ (see Figure \ref{IVcases}(a)). 

\remove{
\begin{figure}[h]
\vspace{-0.1in}
\centering
\includegraphics[scale=0.3]{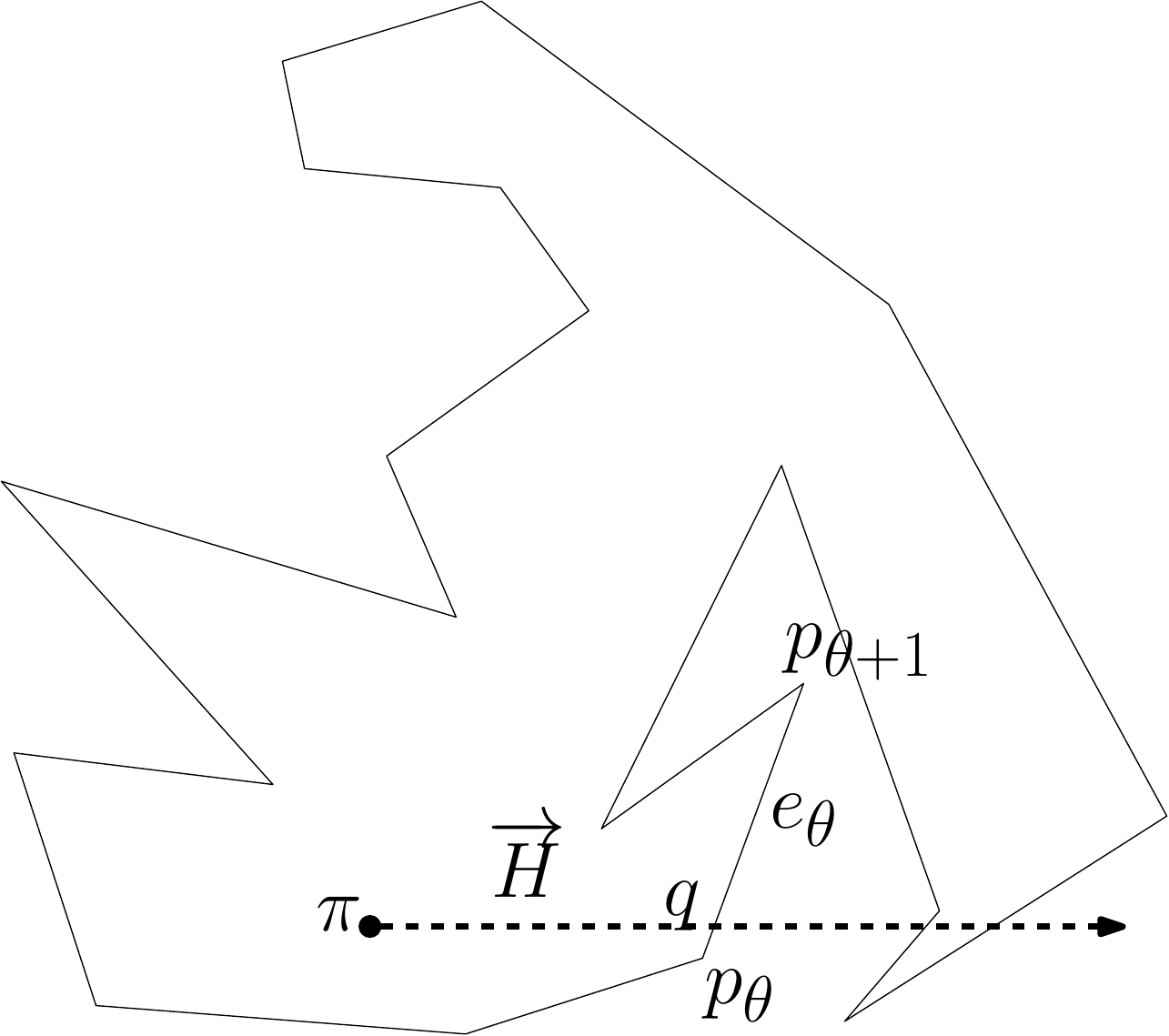}\vspace{-0.1in}
\caption{ Illustration }\vspace{-0.2in} 
\label{polygon}
\end{figure}
}

We visit the vertices of $P$ in counterclockwise order starting 
from $p_{\theta+1}$. After visiting all the vertices of $P$, the
visibility polygon of $\pi$ will be stored in consecutive
locations of the 
input array. Note that, at some point of time during the execution, a vertex may
be visible to $\pi$, but as the algorithm 
proceeds, it may not remain visible. \vspace{-0.1in} 
\begin{observation}
Let $p_i$ and $p_j$ be a pair of visible vertices at an instant of time, 
where $(i-\theta) \bmod{n} < (j-\theta) \bmod{n}$. If both $p_i$ and $p_j$ 
become invisible during the further execution of the algorithm, then $p_j$ 
becomes invisible prior to $p_i$. 
\end{observation}\vspace{-0.1in} 

As in the classical algorithm for computing the visibility polygon \cite{G07}, 
here also we store the  vertices of $P$ {\it visible} to $\pi$ in a stack. At the end of
the execution, the content of the stack indicates the vertices of the
visibility polygon of $\pi$. Due to the constraint on space, here we maintain
the stack in the array $P$ itself. We use three index variables $i$, $k$ and
$\ell$, where $k$ and $\ell$ denote respectively the starting and ending indices
of the stack at an instant of time, and $i$ indicates the index of the current 
vertex of $P$ under processing. We use two more workspaces
$\Phi$ and $\Psi$ that stores $\angle{q\pi
P[\ell]}$ and $\angle{q\pi P[i]}$ respectively, where $\ell$ and $i$ are as
stated above. Note that the content of $\Phi$ can be from $0^o$ to $360^o$. But
$\Psi$ can contain negative angle. This happens when the traversal path along
the boundary of the polygon crosses the ray $\overrightarrow{H} = {\pi q}$ an
even
number of times.

The execution starts from the vertex $P[\theta+1]$. Initially  we set
$k=\ell=\theta+1$, $i=\theta+2$, $\Phi=\Psi=0$. We process each
vertex $P[i]$, $i=\theta+2,\theta+3,\ldots,n,1,2,\ldots,\theta$ in this order.
While processing $P[i]$, we compute $\Psi=\angle{q\pi P[i]}$ as
follows: 
\vspace{-0.1in} 
\begin{description}
\item[$\bullet$] If $P[i-1]\rightarrow P[i]$ is an anticlockwise turn then 
$\Psi=\Psi+\angle{P[i-1] \pi P[i]}$; 
\item[$\bullet$] Otherwise $\Psi=\Psi-\angle{P[i-1] \pi P[i]}$. 
\end{description} 
\vspace{-0.1in} 
Next, we compare $\Phi$ with $\Psi$. Here one of the following three
cases may arise. The appropriate actions in each case is explained below. At the
end of the execution, the visibility polygon is
available in $P[k \ldots,\ell]$.

\begin{description}
\item[Case 1:] $\Phi \le \Psi$ (see Figure
\ref{IVcases}(a)). 
\item[Case 2:] $\Phi > \Psi$ and the polygon makes a
right turn at $P[i]$ (see Figure
\ref{IVcases}(b)).
\item[Case 3:]  $\Phi > \Psi$ and the polygon makes a
left turn at $P[i]$ (see Figure
\ref{IVcases}(c)).
\end{description}

\begin{figure}
\vspace{-0.25in}
\centering
\includegraphics[scale=0.21]{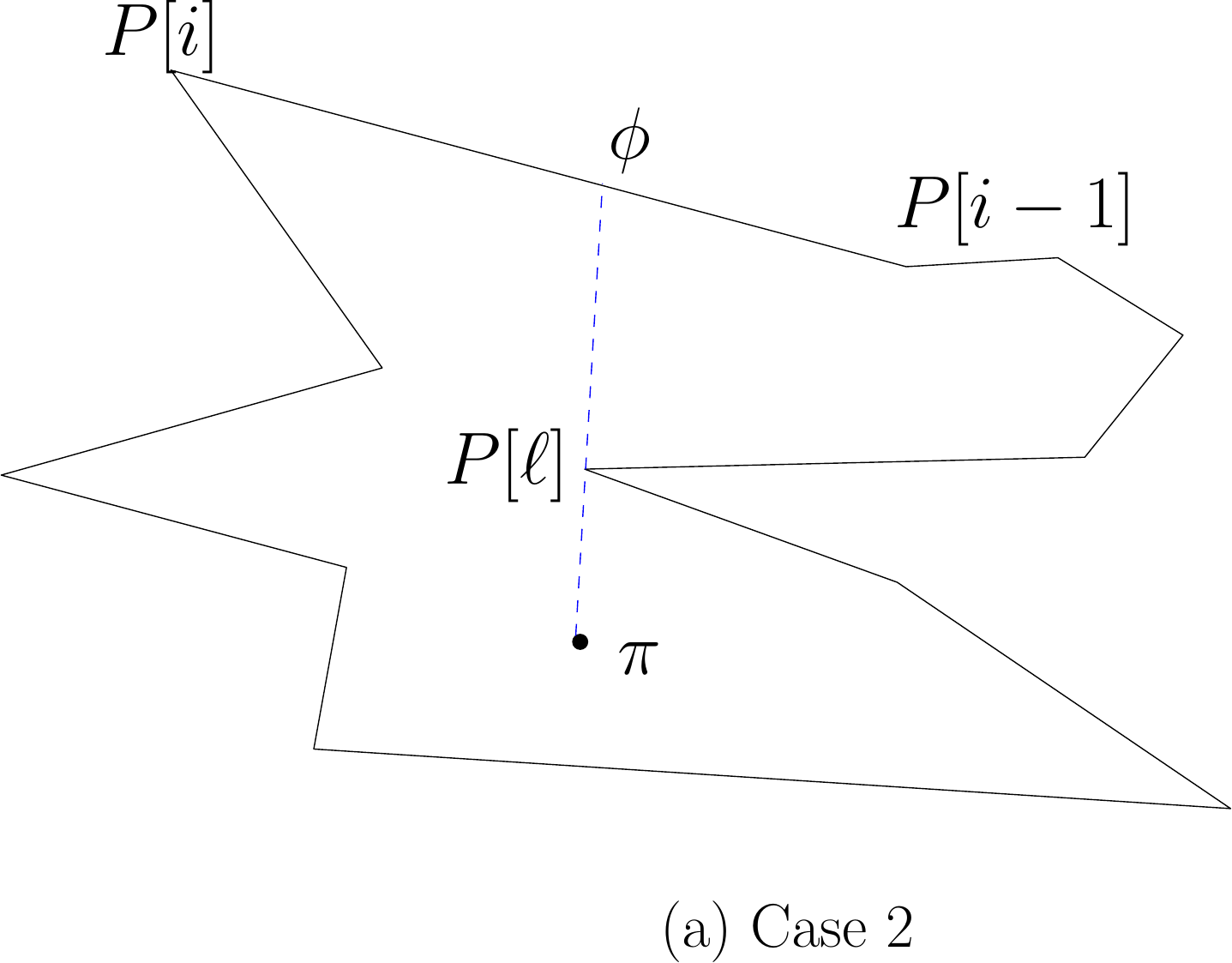}
\includegraphics[scale=0.21]{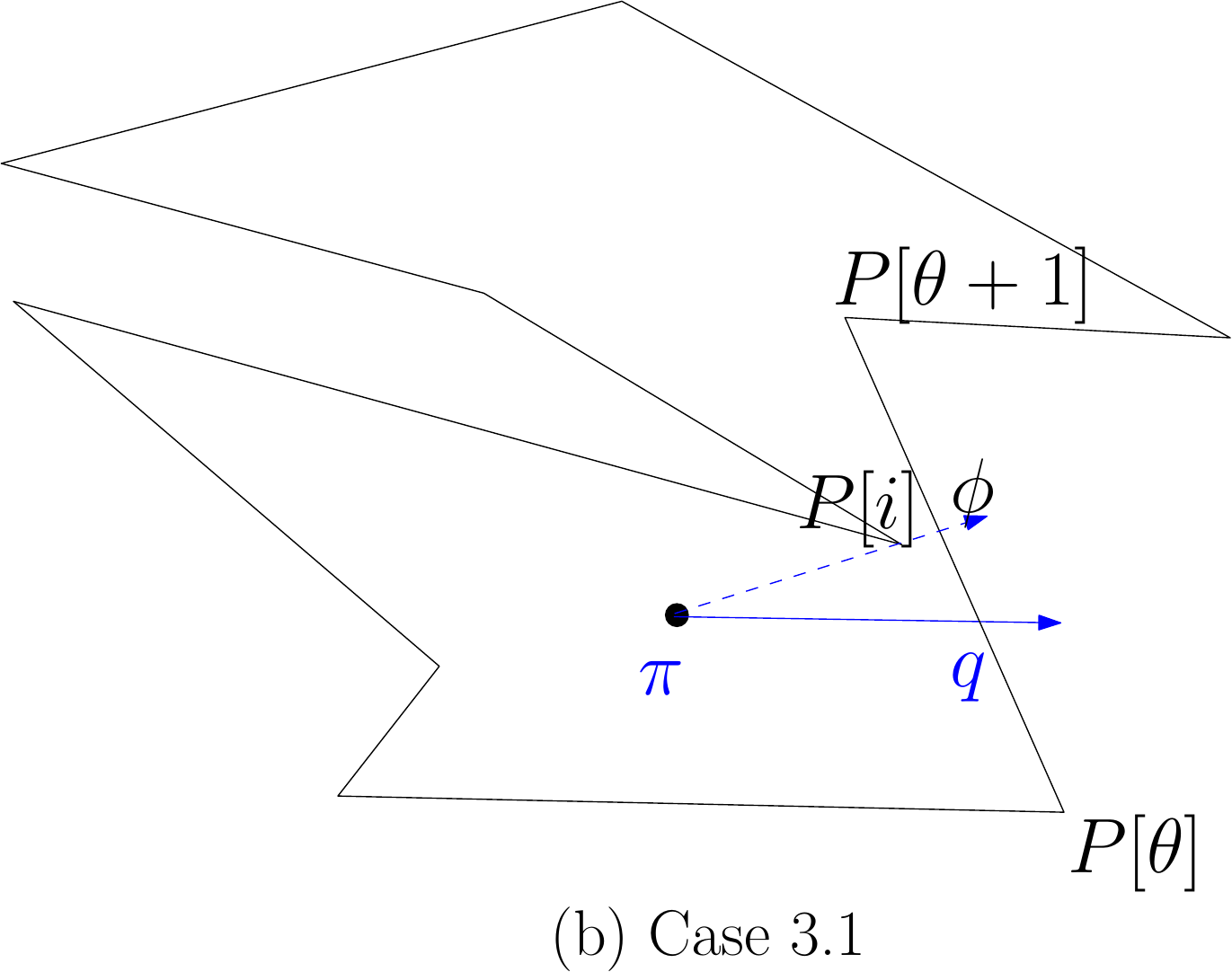}
\includegraphics[scale=0.18]{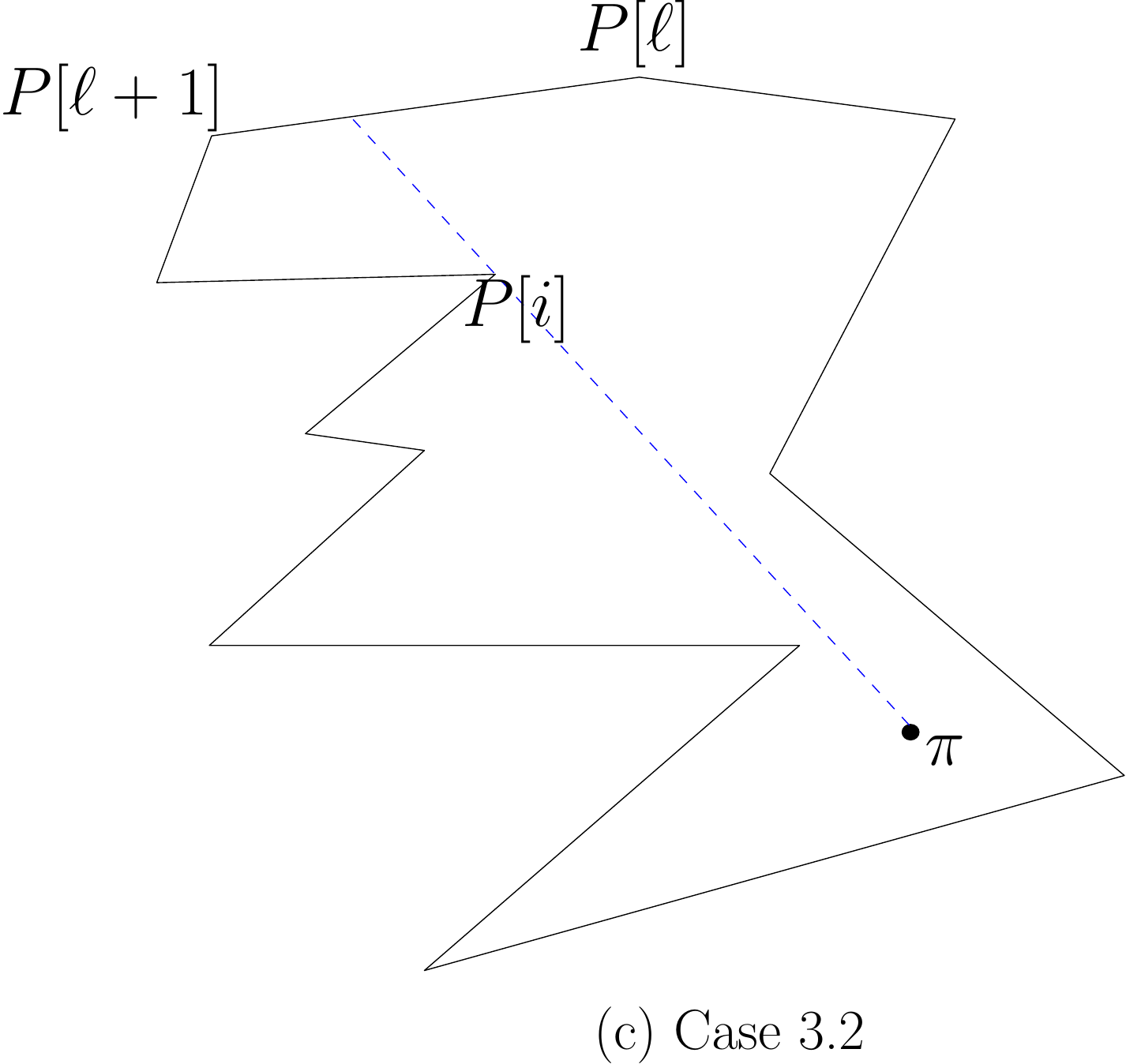}
\includegraphics[scale=0.19]{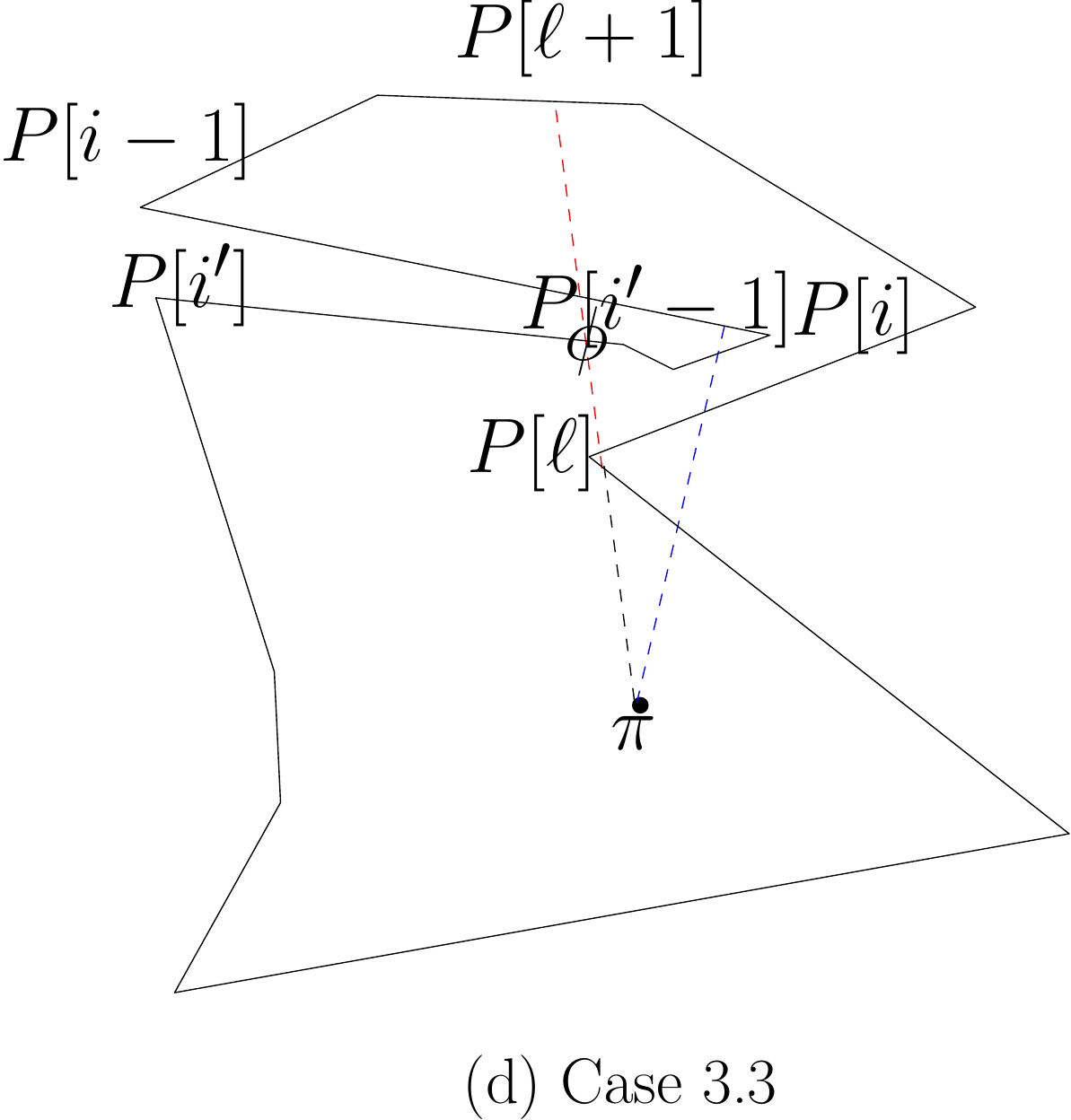}\vspace{-0.1in}
\caption{Different cases that arise during the execution of our in-place
algorithm }
\label{f}
\vspace{-0.2in}
\end{figure}

\vspace{-0.1in}
In Case 1, $P[i]$ is visible from the point $\pi$.  We do the following: (i)
push $P[i]$ in the stack by setting $\ell=(\ell+1) \bmod{n}$ and
placing $P[i]$ in  location
$P[\ell]$, (ii) set $\Phi=\Psi$, and (iii) increment $i$ by setting $i=(i+1)
\bmod{n}$ to process  next vertex.\\
In Case 2,  we ignore $P[i]$ and increment $i$  (by setting $i=(i+1) \bmod{n}$)
until it encounters some vertex satisfying Case 1 (see Figure \ref{f}(a)).\\
In Case 3, we pop elements from the stack by (i) setting $\Phi =
\Phi-\angle{P[\ell] \pi P[\ell-1]}$ and (ii) decreasing  $\ell$ by 1 at each step
until one of the followings hold: \vspace{-0.1in} 
\begin{description}
\item[Case 3.1:] {\bf $\ell$ becomes less than $k$, i.e., the stack becomes
empty.} Here $k$ and $\ell$ are reset to $i$ (see Figure \ref{f}(b)).  
\item[Case 3.2:] {\bf $P[\ell]$ and the current $P[i]$ satisfy $\Phi < \Psi$}
(see Figure \ref{f}(c)). Now, process $P[i]$ as in Case 1.
\item[Case 3.3:]{\bf $P[\ell]$ and $P[i]$ satisfy $\Phi > \Psi$ and the line
segments $[P[i-1],P[i]]$ and $[P[\ell], P[\ell+1]]$ intersect in their interior}
(see Figure \ref{f}(d)). Here we need to proceed (by incrementing $i$ by 1 at
each step) until a vertex $P[i']$ is obtained that satisfies Case 1. Now we
process $P[i']$ as in Case 1.
\end{description} \vspace{-0.1in} 

Note that, in order to check Case 3.3, we need $P[i-1]$ and $P[\ell+1]$;
$P[i-1]$ may be lost during the execution of the algorithm. So 
we need to maintain $P[i-1]$ in a scalar location {\it previous\_vertex}.  
$P[\ell+1]$ is the last deleted vertex from the stack; it is available
since no other vertex is inserted yet in the stack.

\vspace{-0.2in}

\subsubsection{Necessary modification required to obtain the entire visibility
polygon}
In order to compute the entire  visibility polygon, we need to modify the 
execution steps of Case 2 and Case 3 (where $P[i]$ is not visible from $\pi$) as  follows:

In Case 2, after computing the vertex $i$ satisfying Case 1, we create a vertex 
$\phi$ at the point of intersection of the edge $(p_{i-1},p_i)$ and the line 
joining $\pi$ and $P[\ell]$ (see Figure \ref{f}(a)). The point $\phi$ is pushed
in the stack. Next, the
vertex $p_i$ is processed as in Case 1.

In Case 3, while backtracking (popping vertices from stack) by 
decrementing $\ell$ by 1 at each step, we may arrive at one of the above three
situations. 

In Case 3.1, $q$ is a vertex of the visibility polygon. Thus, apart from
resetting
$k$ and $\ell$, we need to insert $q$ in the stack (see Figure \ref{f}(b)). Note
that, during the
entire execution, this case may appear at most once.

In Case 3.2, a new vertex $\phi$ is created at the point of intersection of
the line joining $[\pi,P[i])$ and the edge $(P[\ell],P[\ell+1])$ of $P$
(see Figure \ref{f}(c)). Next, $\phi$ and $P[i]$ are inserted in the stack.

In Case 3.3, after computing $P[i']$, a new vertex $\phi$ is created at the point
of intersection of the line joining $(\pi,P[\ell])$ and the edge $(P[i'-1],
P[i'])$ of $P$ (see Figure \ref{f}(d)). Next, $\phi$ and $P[i']$ are inserted in
the stack.

The pseudo-code of Algorithm~\ref{Vis} is given in the appendix. The correctness
of the algorithm follows from Lemmas 3-7 of \cite{L83} and Lemmas \ref{lll}
stated below.
\vspace{-0.1in}

\begin{lemma} \label{lll}
The inplace maintenance of the stack in the same array $P$ does not erase the 
polygonal vertices prior to its processing.
\end{lemma}
\vspace{-0.2in}

\begin{proof}
Consider a {\it push} operation for maintaining the visibility polygonal vertices 
during the execution. Such a vertex may be either (i) a vertex of $P$ or (ii) a 
point on an edge of $P$. When a polygonal vertex is pushed in the stack (Case
(i)), $i$ is immediately incremented. Thus, it does not erase any unprocessed
polygonal vertex. In Case (ii), we need to mention that each pair of consecutive
edges of the stack defines an edge of the visibility polygon \cite{L83}. While 
processing $P[i]$, if a point $\phi$ on the visibility polygonal edge $(\alpha,
\beta)$ is created, which is not a polygonal vertex, then both $\alpha$ and
$\beta$ are present in the stack. We pop $\alpha$ and push $\phi$ in the stack.
Thus the stack does not overlap unprocessed vertices. \qed
\end{proof}

\vspace{-0.1in}

Apart from the array $P$, we need at most a constant number of scalar locations
to run the algorithm. Its time complexity depends on the number of changes in
the value of $i$ and $\ell$. Since $i$ is never decremented, $i$ is modified $n$
times. $\ell$ is also modified $O(n)$ times since it is incremented at most $n$
times, and the number of decrements of $\ell$ is bounded by the number of its
increments. If there is any reset in $k$ during the execution, then $\ell$ is
also changed. Thus changes in $k$ does not affect the complexity of the
algorithm. Thus, we have the following result:
\vspace{-0.1in}

\begin{theorem}
The time complexity of the algorithm \IV\ is $O(n)$, and it uses $O(1)$ extra
work-space. 
\end{theorem}

%%%%%%%%%%%%%%%%%%%%%%%%%%%%%%%%%%%%%%%%%%%%%%%%%%%%%%%%%%%

\vspace{-0.3in}
\subsection{\RV}
\vspace{-0.1in}
Here, we assume that the vertices of the input polygon are given in a
read-only array $P$ in counterclockwise order, and show that the visibility
polygon for a point $\pi$ inside $P$ can be computed in $O(n)$ time using
$O(\sqrt{n})$ extra spaces.

As in the earlier section, here also we draw a horizontal ray
$\overrightarrow{H}$ from the point $\pi$ towards right that meets the edge
$e_\theta = (p_\theta,p_{\theta+1})$. We first partition the polygon into
$\lceil\sqrt{n}\rceil$ polygonal chains (polychains); each containing $\sqrt{n}$
vertices except the last one, which may contain fewer number of vertices. The
first vertex of the first chain is $p_{\theta+1}$. The $j$-th polychain will be
referred to as $P_j$. We start processing from $p_{\theta+1} \in P_1$, and
process the polychains in counterclockwise order. While processing $P_j$, the
vertices of $P_j$ are also processed in counterclockwise manner.
\vspace{-0.2in}
\begin{definition}
A polychain $P_j$ is called  {\em processed} if all its vertices are
processed. 
\end{definition} \vspace{-0.1in}
We maintain two arrays of integers: $S$ of size $\lfloor\sqrt{n}\rfloor$, 
and $R$ of size $2 \times \lceil\sqrt{n}\rceil$.

Let us assume that we have processed the chains $P_1,\cdots,P_{j-1}$ and next we
want to process $P_j$. The array $S$ contains the indices of the vertices in
$P_j$. At an instant of time, if any part of the polychain $P_j$ is visible
from $\pi$, then $R[1,j]$ and $R[2,j]$ stores the indices of two vertices
of $P$ that blocks the visibility of $P_j$ from its left and right sides at 
that instant of time. If the first (resp. last) vertex of $P_j$ is visible to
$\pi$, then $R[1,j]$ (resp. $R[2,j]$) stores that vertex itself. A zero entry
in $R[1,j]$ and $R[2,j]$ indicate that $P_j$ is entirely not visible to $\pi$.
Note that, if more than one part of $P_j$ is visible from $\pi$ at an instant of
time, then that information is not stored in the array $R$. From now onwards, by
the term that a vertex is visible/invisible, we mean that it is
visible/invisible to $\pi$. The following two structural lemmas are crucial.
\vspace{-0.1in}
\begin{lemma}
\label{l2}
After the processing of $P_j$, let some vertices in the chain $P_k$ ($k < j$)
remains visible; the minimum and maximum indices of visible vertices in $P_k$ be
$f_k$ and $l_k$, respectively. Now, if there exists any invisible vertex
$p_\beta \in P_k$ with $f_k < \beta < l_k$, then the visibility of $p_\beta$ 
can only be obstructed by an edge of the chain $P_k$ itself. 
\end{lemma}
\vspace{-0.2in}
\begin{proof}
If the visibility of $p_\beta$ is obstructed by some edge of an unvisited
polychain, then it is not yet identified. So, we need to consider the case where
$p_\beta$ is obstructed by some edge of a {\it visited} polychain $P_\gamma$,
where $1< \gamma \leq k$ or $k < \gamma\leq j$. In the former case, prior to
obstructing $p_\beta$, it must have obstructed $p_{f_k}$. In the latter case, 
prior to obstructing $p_\beta$, it must have obstructed $p_{l_k}$. In both the
cases we have contradiction since both $p_{f_k}$ and $p_{l_k}$ are visible to
$\pi$. \qed
\end{proof}

\vspace{-0.2in} 
\begin{lemma} \label{l1}
During the processing of $P_j$, if it is observed that some visible vertices of
$P_k$ ($k < j$) becomes invisible, then all the  vertices of $P_{j-1},$ 
$P_{j-2}, \ldots, P_{k+1}$ (if any) becomes invisible. Moreover, if a vertex 
$p_\alpha \in P_j$ becomes invisible, then all the vertices having index greater
than $\alpha$ in $P_j$ are invisible. 
\end{lemma}\vspace{-0.1in} 
\remove{
\begin{proof}
Similar to the proof of Lemma \ref{l2}. \qed
\end{proof}
}
Our algorithm consists of two passes. In the first pass, we compute $f_j$ and
$l_j$ for all the polychains $P_j$ in anticlockwise order, and set $R[1,j]$ and
$R[2,j]$ as described above. In the second pass, we consider each $P_j$ and
print the visible portions from $\pi$ (if any). If $R[1,j], R[2,j] \neq 0$, then
there exists at least one part of $P_j$ that is visible from $\pi$. 
\vspace{-0.2in}
\subsubsection{Pass 1:}
In this pass, we process $P_j$ as in Subsection \ref{IV}. We copy the indices of
the vertices of $P_j$ in array $S$, and process them in counterclockwise order.
As in Section \ref{IV}, during the execution, the visible portion of $P_j$ is
stored in the stack maintained at the beginning of $S$. An index variable $\ell$
is used as the top pointer of the stack. We also use a variable $\chi$ that
contains the index of the most recently visited polychain which is
completely/partially visible at the current instant of time. At the beginning of
processing $P_j$, $\chi$ is initialized with $j-1$. 

While processing a vertex $p \in P_j$, here also three cases may arise. The 
processing of different cases are same as in Subsection \ref{IV} except Case 2
and Case 3.1.  

In Case 2, if the last vertex $p$ of $P_j$ is not visible (i.e. $\angle{q\pi
S[\ell]} > \angle{q\pi p}$), then we store $S[\ell]$ in a temporary variable
$\sigma$, We set $R[2,j] = R[1,j+1] = \sigma$, and copy $P_{j+1}$ in $S$ for the
processing.  

In Case 3.1, the first task is to put the index of $p$ in $R[1,j]$. Next, we
check whether the vertex $p$ blocks the visibility of some already processed
polychains by considering them in clockwise order starting from $P_\chi$. By
Lemma \ref{l1}, a polychain $P_k$ is considered if all the polychains
$\{P_\gamma, k < \gamma  < j\}$ become invisible. While considering $P_k$, we
may have the following three situations. Let the vertex $p'$ be the
clockwise neighbor of the vertex $p\in P_j$, $R[1,k]=\alpha$ and $R[2,k]=\beta$.
\begin{description}
\item[Case 3.1.1 -] {\bf The edge $(p',p)$ does not intersect the lines $[\pi, 
p_\alpha)$ and $[\pi,p_\beta)$:} Here the entire visible portion of the
polychain $P_k$ remains visible, and we need not have to test the other 
visited polychains $P_\gamma$, $\gamma < k$. So, we push $p$ on to the
stack maintained in $S$, and consider the vertex next to $p$ in the polygon $P$
for processing.

\item[Case 3.1.2 -] {\bf The edge $(p',p)$ of $P$ intersects the line $[\pi,
p_\beta)$ but does not intersect $[\pi,p_\alpha)$: } Here, we replace $\beta$ 
by the index of $p$ in $R[2,k]$. Next, $p$ is pushed in $S$ and the vertex
next to $p$ is considered for processing.

\item[Case 3.1.3 -] {\bf The edge $(p',p)$ of $P$ intersects both the lines
$[\pi,p_\alpha)$ and $[\pi,p_\beta)$: } Here the entire polychain $P_k$
becomes invisible. We set $R[1,k]=R[2,k]=0$, and consider the next visible
polychain in clockwise order. The process continues until we arrive at either
Case 3.1.1 or Case 3.1.2.
\end{description}\vspace{-0.1in}
At the end of processing $P_j$(if special Case 2, as mentioned before,
doesn't arise), we put the index of the last vertex of
$P_j$ in $R[2,j]$.
\vspace{-0.2in}
\subsubsection{Pass 2:}
In this pass, we consider each polychain $P_j$ for printing its visible
portion from $\pi$. We start from $P_1$, and proceed  in counterclockwise
order. If $R[1,j], R[2,j] \neq 0$ for a polychain $P_j$, then it is
fully/partially visible from $\pi$. We compute the leftmost and rightmost 
points of $P_j$, say $f_j$ and $l_j$ that are visible from $\pi$ as follows. 
\vspace{-0.1in} 
\begin{itemize}
\item[] If $R[1,j]$ points to the leftmost point of $P_j$, then $f_j = R[1,j]$.
Otherwise, we identify the edge $e$ that is hit by the ray
$\overrightarrow{\pi p_{\alpha'}}$ first, where $\alpha'$=$R[1,j]$. We
compute $\phi$ = point of intersection of $e$ and $\overrightarrow{\pi
p_{\alpha'}}$. 
\item[] Similarly, if $R[2,j]$ points to the rightmost point of $P_i$, then
$l_j=R[2,j]$. Otherwise, we identify the edge $e$ that is hit by the ray
$\overrightarrow{\pi p_{\beta'}}$ first, where $\beta'=R[2,j]$. We
compute $\psi$ = point of intersection of $e$ and $\overrightarrow{\pi
p_{\beta'}}$. 
\end{itemize}\vspace{-0.1in}
Next we copy the indices of all the vertices of $P_j$ between $\phi$ and
$\psi$ along with $\phi$ and $\psi$ in the array $S$ in order. By Lemma
\ref{l1}, if there is any invisible vertex in $S$, it is blocked by some
edge in $P_j$. So, we execute \IV\ on the array $S$. At the end of processing
$P_j$, the content of the stack is printed. 
\vspace{-0.1in}
\begin{theorem}
The algorithm \RV\ correctly computes the visibility polygon of $P$ from the 
point $\pi$. It needs $O(n)$ time and $O(\sqrt{n})$ work-space.  
\end{theorem}\vspace{-0.1in}
\begin{proof}
The correctness of the algorithm \RV\ follows from the fact that for each 
polychain $P_j$, if its visibility is blocked from any/both side(s), then
blocking
vertices are correctly computed as mentioned in the correctness proof of
the algorithm \IV\, and are stored in the array $R$. In Pass 2, the first
and last visible vertices
of $P_j$ can be correctly computed from the content of $R[1,j]$ and
$R[2,j]$. The correctness of Pass 2 follows from Lemma \ref{l2}. We now analyze
the complexity results of the algorithm.

While executing Pass 1
on the polychain $P_j$, copying the polychain $P_j$ in $S$ needs
$O(\sqrt{n})$ time. During the processing of Pass 1, each vertex in $P_j$ is
inserted in and deleted from stack at most once. In the entire Pass 1, if a
polychain becomes completely invisible, i.e., $R[1,j],R[2,j]$ are set to 0, it
never becomes visible. Note that, during the processing of each vertex in $P_j$,
the visibility of at most one polychain is reduced by changing its $R[1,j]$
field. Thus, the amortized time complexity for processing each vertex in Pass 1
is $O(1)$. While executing Pass 2 for each $P_i$, we first spend $O(\sqrt{n})$
time for computing the first and last visible vertices $\phi$ and $\psi$. The
copying of the visible portion of $P_i$ in $S$, and executing \IV\ on $S$ needs
another $O(\sqrt{n})$ time in the worst case. Since we have
$\lceil\sqrt{n}\rceil$ polychains, the result follows. \qed
\end{proof}

%%%%%%%%%%%%%%%%%%%%%%%%%%% WEAK VISIBILITY  %%%%%%%%%%%%%%%%%%%%%%%%%%%%%%
%%%%%%%%%%%%%%%%%%%%%%%%%%%%%%%%%%%%%%%%%%%%%%%%%%%%%%%%%%%%%%%%%%%%%%%%%%%
\vspace{-0.3in}
\section{Weak Visibility Polygon of an edge} \label{WV}
\begin{figure}[b]
\vspace{-0.2in}
\centering
\includegraphics[scale=0.2]{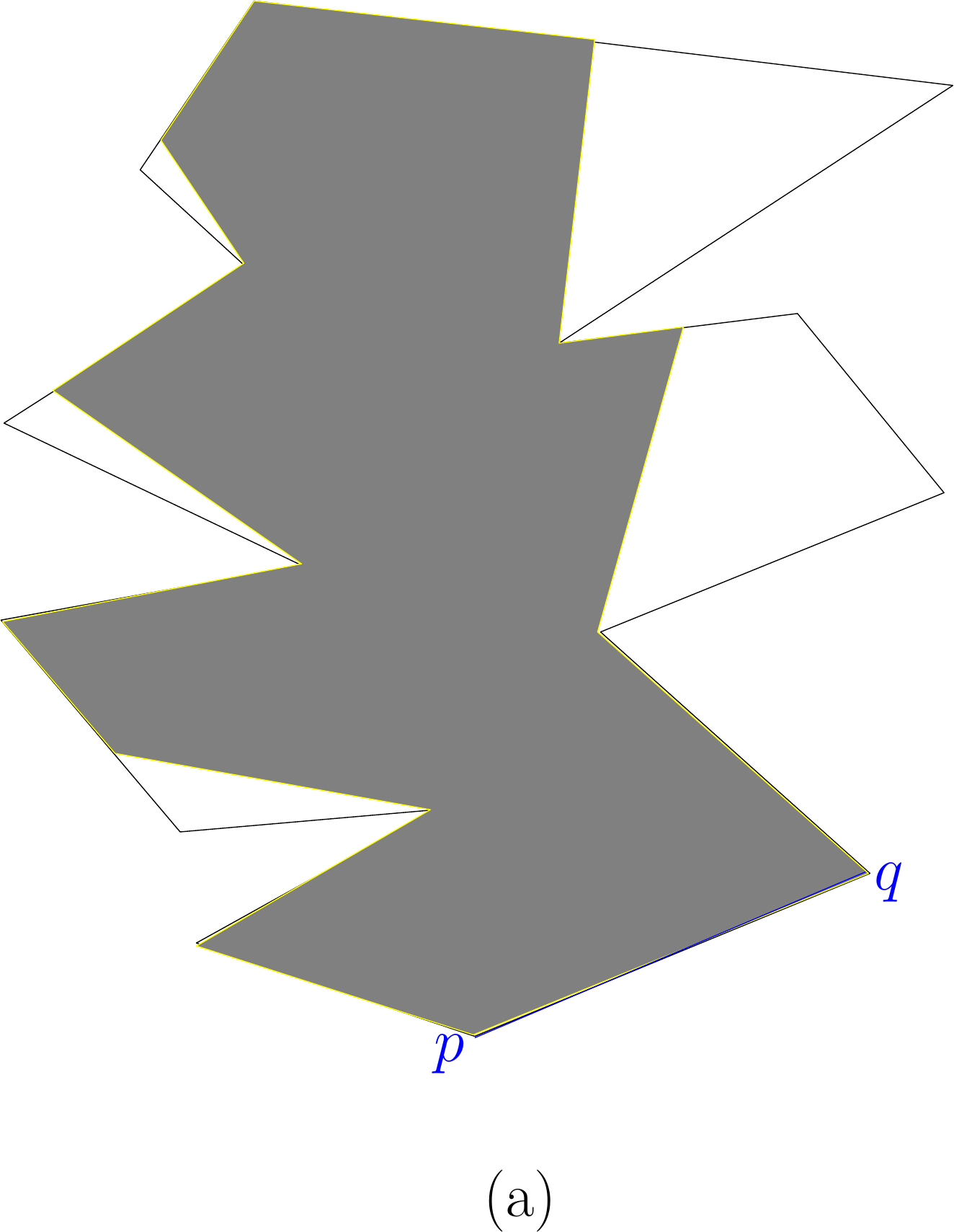} \hspace{0.2in}
\includegraphics[scale=0.2]{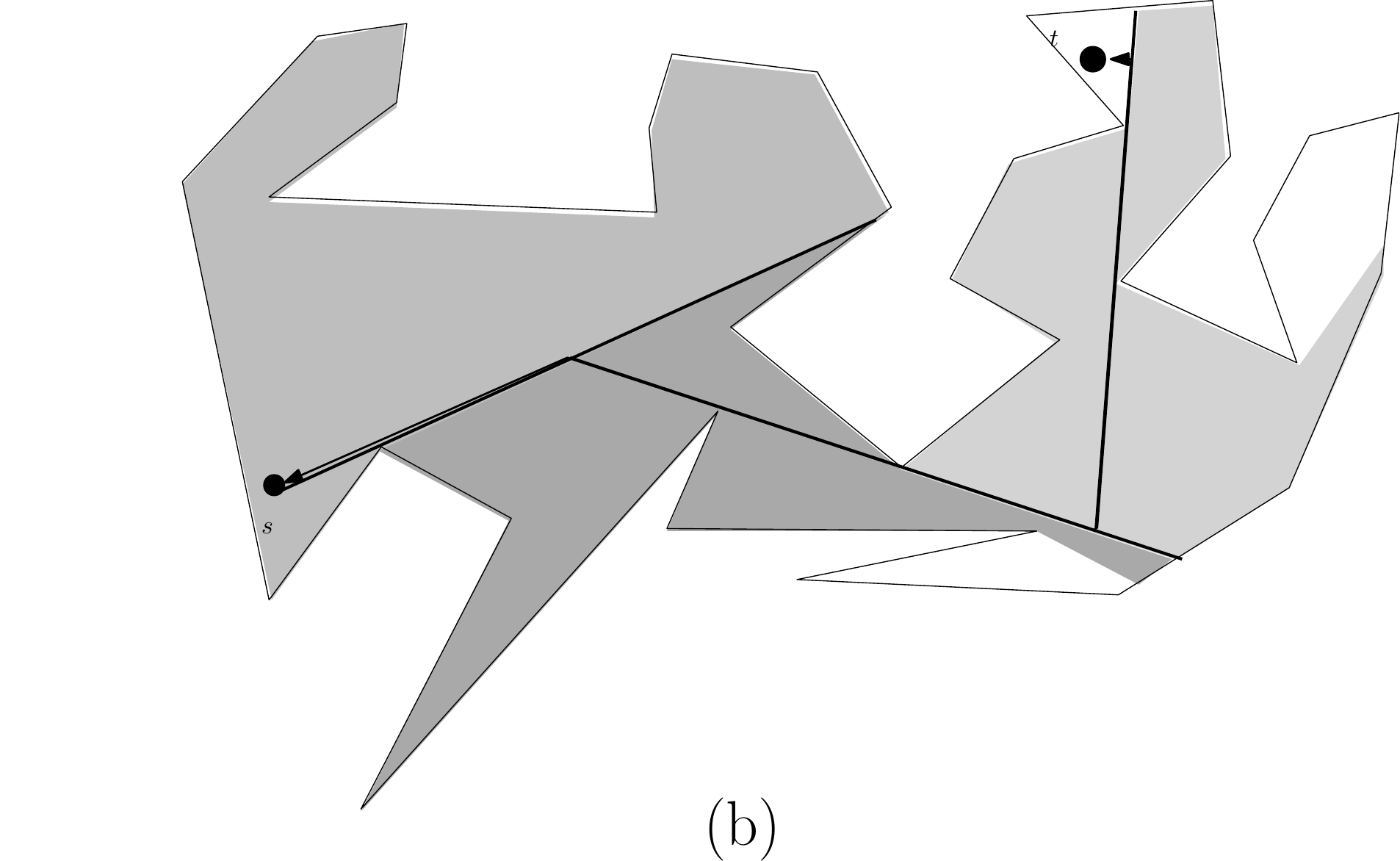}
\vspace{-0.1in}
\caption{Demonstration of (a) Weak visibility polygon, (b) minimum link
path}\vspace{-0.1in}  
\label{fig:wvp}
\end{figure}

\vspace{-0.1in}
Given a polygon $P$ and a line segment $\ell=[p,q]$ in $P$, the weak visibility
polygon of $\ell$, denoted by $WVP(\ell)$, is a simple polygonal region $R$ such
that each point in $R$ is visible from at least one point of $\ell$. As in the
earlier section, here also we will assume that the vertices $\{p_1,p_2,\ldots,
p_n\}$ of the polygon $P$ are given in a read-only array, called $P$, in
anticlockwise order. We will propose an algorithm for computing the
weak-visibility polygon of an edge $e=(p_i,p_{i+1})$ of $P$, denoted by
$WVP(e)$ (see Figure \ref{fig:wvp}).

Before presenting the algorithm for computing $WVP(e)$, let us consider first
the following two simpler problems. 
\vspace{-0.1in}
\begin{description}
\item[$\mathbf{\LNV(e,e')}$:] Given two edges $e$ and $e'$ of a polygon $P$,
compute the portion of $e'$, from its left endpoint,   which is not
weakly-visible from $e$. 

\item[$\mathbf{\RNV(e,e')}$:] Given two edges $e$ and $e'$ of a polygon $P$,
compute the portion of $e'$, from its right endpoint,  which is not
weakly-visible from $e$.  
\end{description}
\vspace{-0.1in}

We will explain the method of solving the problem $\LNV(e,e')$. $\RNV(e,e')$
will be symmetric to that. Once we have the solution of these two problems, we
will be able to compute the portion of $e'$ which is weakly visible from  $e$.
In order to compute $WVP(e)$, we need to compute the weak-visible portion of all
the edges $e' \neq e$ of the polygon $P$.
                                              
\vspace{-0.2in}
\subsection{$\LNV(e,e')$} \label{LNV}
\vspace{-0.1in}
Let $e=[p_i,p_{i+1}]$  and $e'=[p_j,p_{j+1}]$, $j > i$. We first join
$[p_{i+1},p_{j+1}]$ (say $L$). We will use $\theta$ and $\theta'$ to denote the
end-points of $L$ on $e$ and $e'$ respectively. At the end of the
execution, it returns $\Phi$ = left non-visible portion of $e'$ from $e$. 
Initially  $\theta=p_{i+1}$ and $\theta'=p_{j+1}$. The algorithm consists
of three passes. 

\begin{figure}[t]
\centering
\includegraphics[scale=0.18]{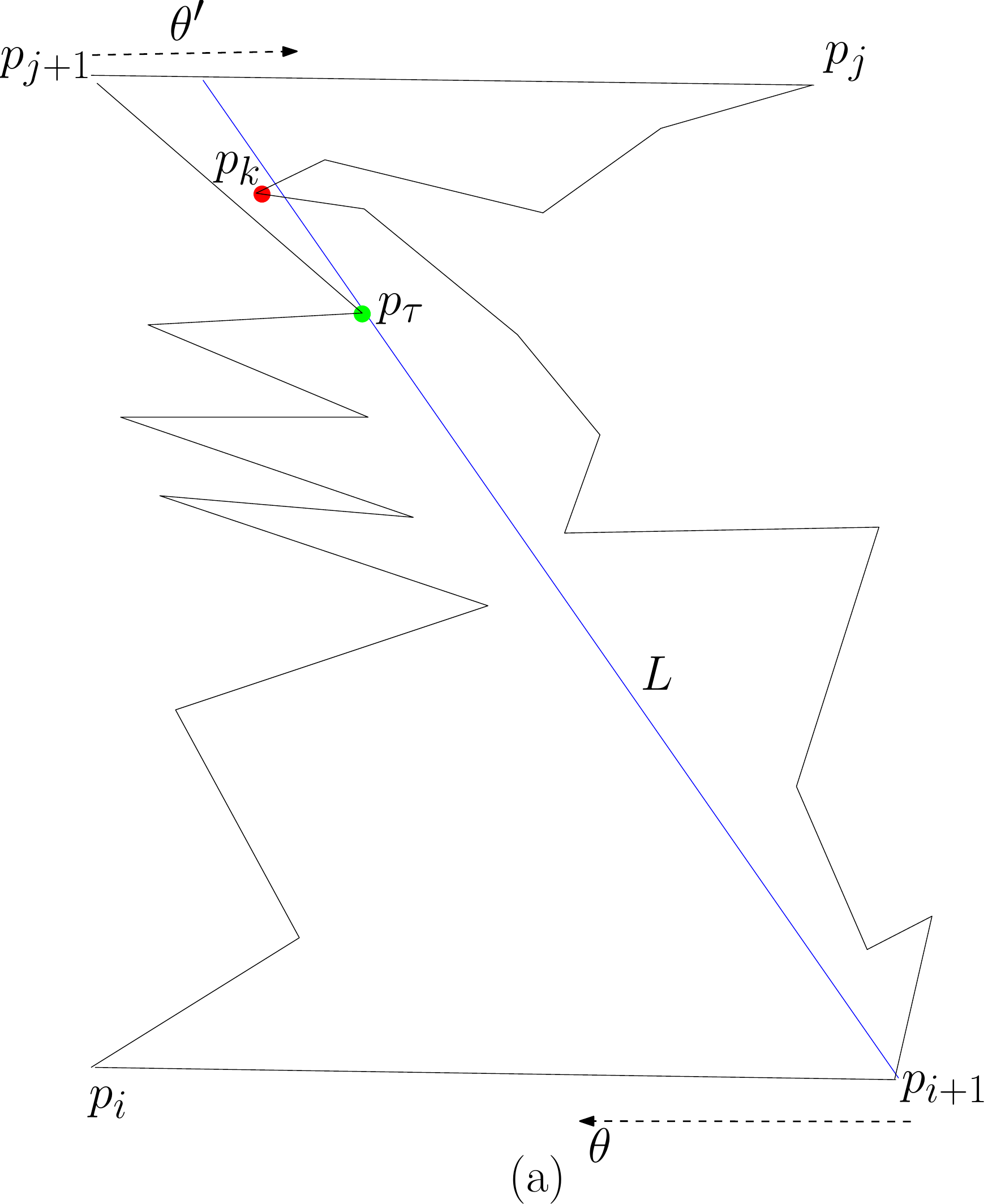}
\includegraphics[scale=0.18]{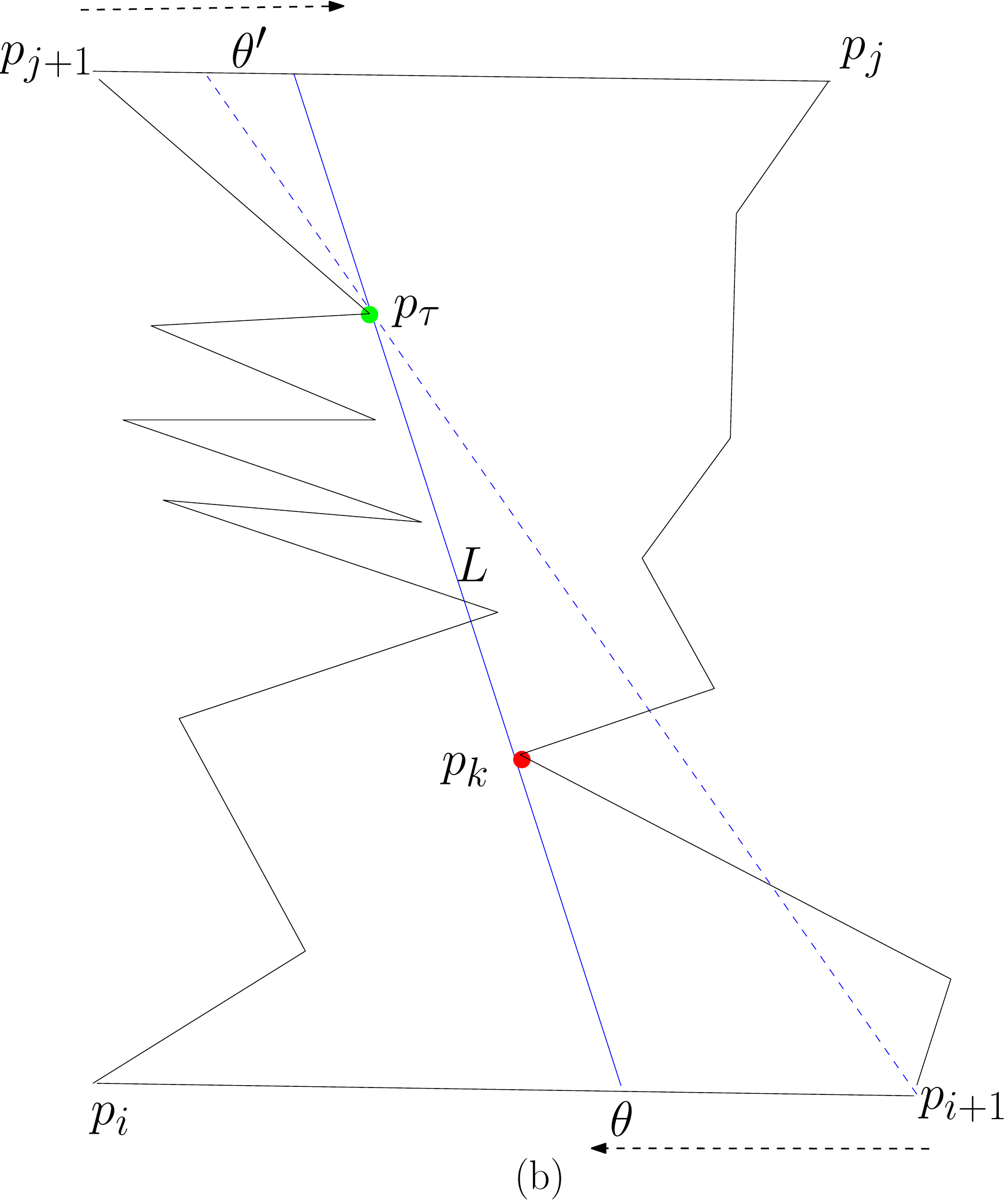}
\includegraphics[scale=0.18]{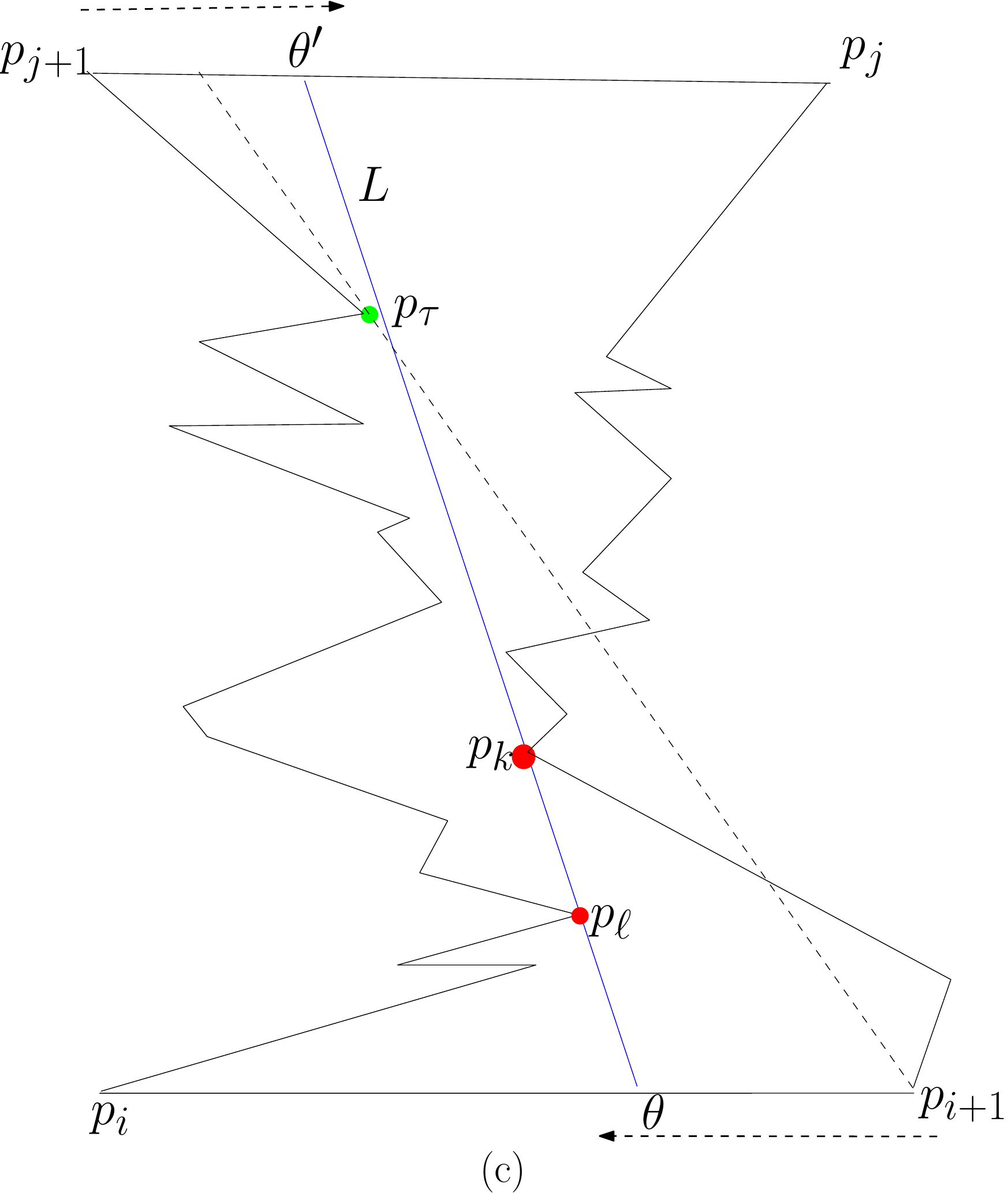}\\
\includegraphics[scale=0.18]{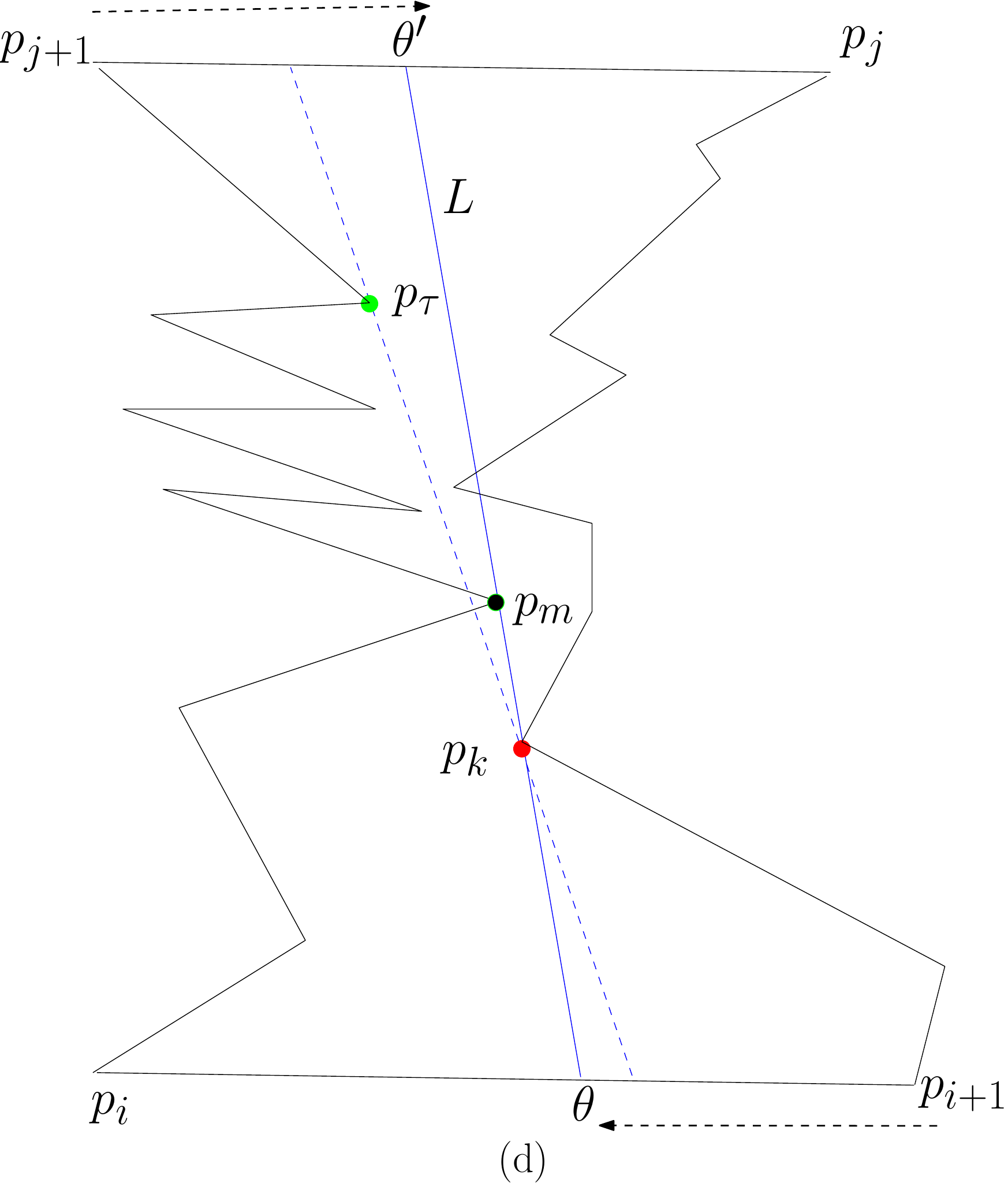}
\includegraphics[scale=0.18]{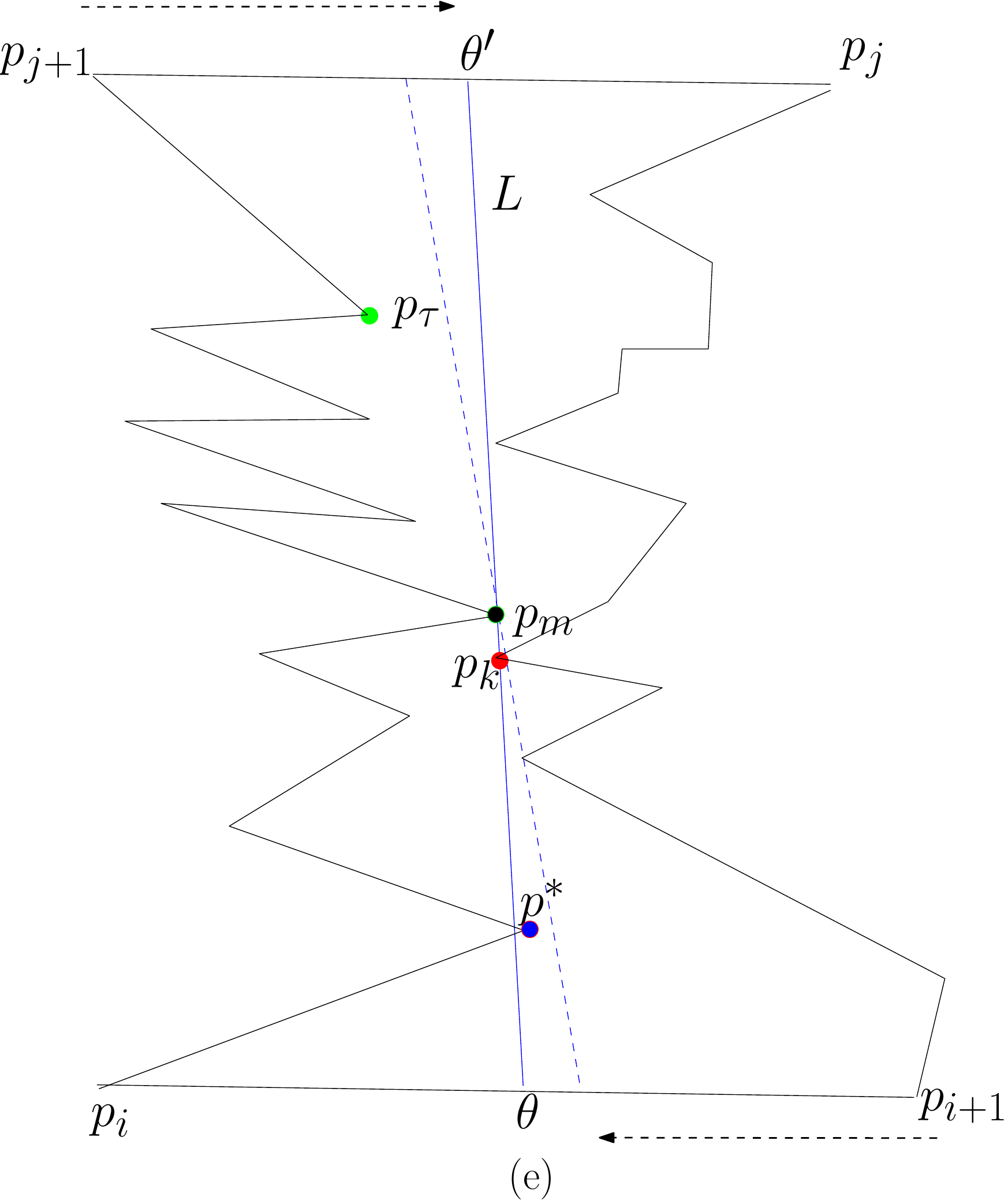}
\includegraphics[scale=0.18]{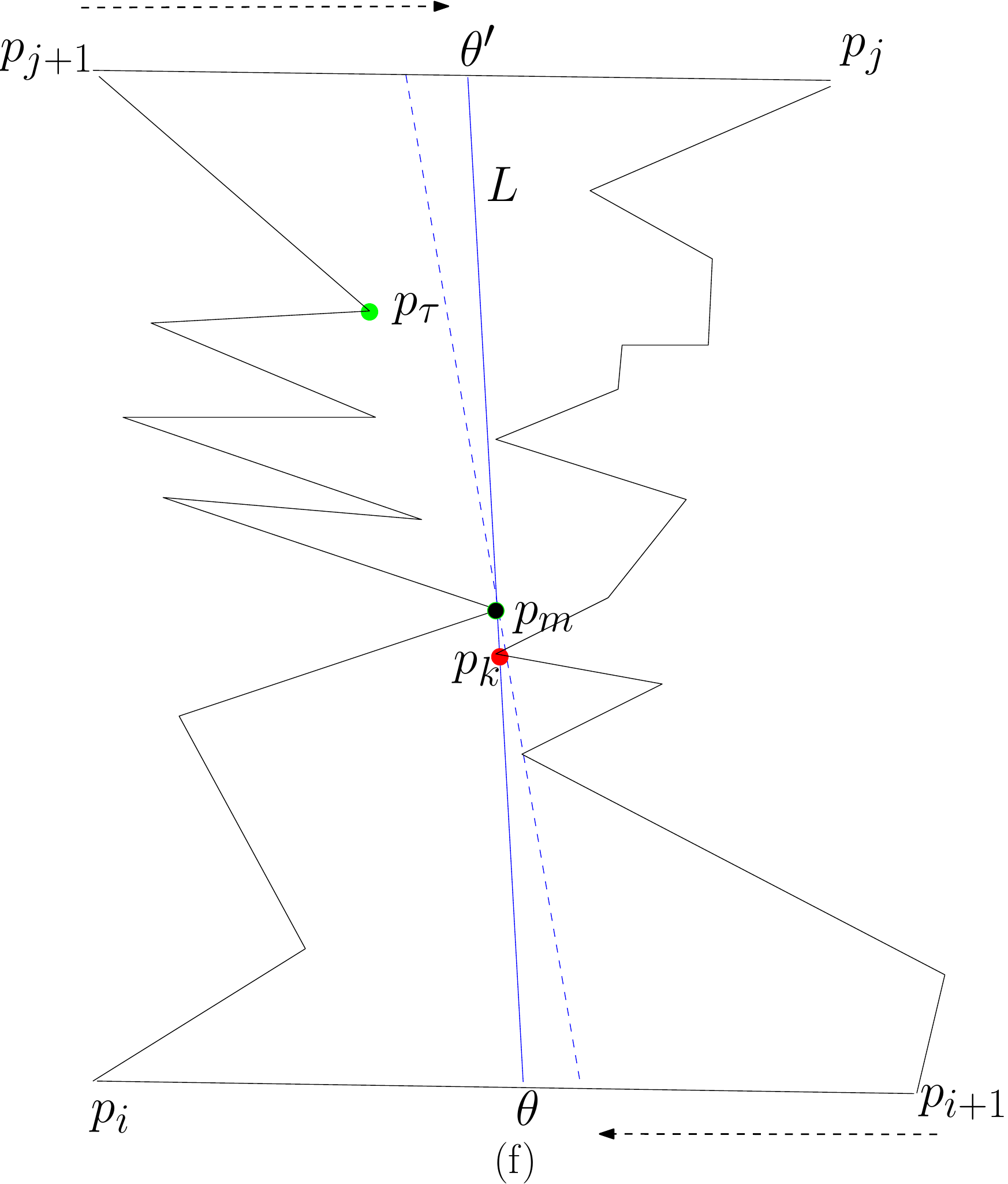}
\caption{Algorithm \LNV}\vspace{-0.2in}
\label{fig:lnvx}
\end{figure}

{\bf Pass-1:} In this pass, polychain $\Pi=[p_{j+1},p_{j+2},
\ldots,p_n,p_1,p_2,\ldots, p_i]$
is traversed in a counterclockwise manner starting from $p_{j+1}$. If a vertex 
$p \in \Pi$ is observed which is to the right of $L$ in its present position, 
then $\theta'$ is moved to the point of intersection of the line containing $e'$ 
and the line $(p_{i+1},p)$; $\theta$ remains fixed at $p_{i+1}$. If $\theta'$ 
is outside $e'$, then $e'$ is not visible from $e$, and the 
procedure returns $\Phi=e'$. Otherwise, $L$ is defined by $p_{i+1}$
and a vertex $p_\tau \in \Pi$ (see Figure \ref{fig:lnvx}(a)). A temporary
variable $\tau$ is used to remember $p_\tau$. 
\remove{
\begin{figure}
\centering
\includegraphics[scale=0.30]{lnv1.pdf}
\includegraphics[scale=0.30]{lnv1a.pdf}
\caption{Algorithm \LNV: Pass-1}
\label{fig:lnv}
\end{figure}
}

{\bf Pass-2:}
In this pass, we simultaneously traverse  $\Pi$ from $p_i$ in clockwise
order up to the vertex $p_\tau$, and $\Pi'$ from $p_{i+1}$ to $p_j$ in
counterclockwise order. The method of traversal is
explained in {\it Process-1} and {\it Process-2}, stated below. We use three
index variables $k$, $\ell$ and $m$. Initially, we set $k=i+1$,
$\ell=i$ and $m=\tau$.

{\bf Process-1:} We traverse $\Pi$ in anticlockwise direction using the index
variable $k$. At each move one of the following events may be observed:
\vspace{-0.1in}
\begin{itemize}
\item [$\bullet$] $k= j$. Pass 2 stops, and it returns
$\Phi=[p_{j+1},\theta']$. 
\item [$\bullet$] $k\neq j$ and the edge $(p_{k-1},p_k)$ does not intersect $L$.
Here $k$ is incremented ($\bmod {n}$) to process the next vertex of $\Pi$.
\item [$\bullet$] $k\neq j$ and the edge $(p_{k-1},p_k)$ intersects $L$ above
$p_m$. Here, $e'$ is not weakly visible to $e$ (see Figure \ref{fig:lnvx}(a)), 
The procedure returns $\Phi=e'$. 
\item [$\bullet$] $k\neq j$ and the edge $(p_{k-1},p_k)$ intersects $L$
below $p_m$. Here, we update $L$ with the line joining $(p_m,p_k)$, and update
$\theta$ (resp. $\theta'$) by the point of intersection of $L$ and $e$ (resp.
$e'$) (see Figure \ref{fig:lnvx}(b)). Next we execute {\it Process-2}. 
\end{itemize}\vspace{-0.1in}

{\bf Process-2:} We traverse $\Pi'$ in clockwise direction using the index
variable $\ell$. At each move the following events may happen:
 \vspace{-0.1in}

\begin{itemize}
\item [$\bullet$] the edge $(p_{\ell+1},p_\ell)$ does not intersect $L$. 
Here, $\ell$ is decremented ($\bmod {n}$) to process the next vertex of $\Pi'$.
\item [$\bullet$] the edge $(p_{\ell+1},p_\ell)$ intersects $L$ below 
$p_k$. Here, $e'$ is not weakly visible to $e$ (see Figure \ref{fig:lnvx}(c)). 
The procedure returns $\Phi=e'$.
\item [$\bullet$] the edge $(p_{\ell+1},p_\ell)$ intersects $L$ above 
$p_k$. Here, we update the $L$ by the line joining $p_\ell$ and $p_k$.
We also set $\theta'=\pi$ and $m=\ell$, and then switch to executing the {\it 
Process-1}. 
\end{itemize}

 \vspace{-0.1in}

{\bf Pass-3:}
Note that after executing {\it Process 2}, $L$ may intersect $\Pi$ below $p_k$
(see Figure \ref{fig:lnvx}(e)). Thus the weak-visibility of $e'$ from $e$ may be
lost. So, after the successful finish of Pass-2, we execute Pass-3 to check for
possible intersection of $L$ and the edges of $\Pi'$. If such an intersection is
observed, $e'$ is not weakly visible from $e$; otherwise, the final position of
$\theta'$ determines the non-visible portion $\Phi=[p_{j+1}, \theta']$ of $e'$ 
from $e$ (see Figure \ref{fig:lnvx}(f)).

\vspace{-0.1in}

\begin{lemma} \label{l4}
Algorithm \LNV\ correctly computes the invisible portion of $e'$ from
its left
end-point, and it needs $O(n)$ time and $O(1)$
space.
\end{lemma}
\vspace{-0.2in}

\begin{proof}
Initially, we draw the line segment $L=[p_{i+1},p_{j+1}]$. Now,
three  cases may arise: (i) the entire segment $L$ lies inside the
polygon, (ii)  the entire segment $L$ lies outside the polygon, and (iii)
$L$ spans both inside and outside the polygon. In case (i), $e'$ starting
from $p_{j+1}$  is visible. In other two
cases,  there may exist situations where $e'$ is/is not weakly visible from
$e$. We now show that in both cases our algorithm correctly computes the
left non-visible portion of $e'$.

In Pass-1, we compute the invisible portion of $e'$ from left by traversing the
left chain $\Pi=[p_{j+1} \sim p_i]$ (assuming the right chain is absent). 
At each step of traversal, if the current edge obstructs $L$, then $L$ is
shifted
towards right along $e'$, and in the further scanning of the chain $L$
is never shifted towards left. 
 
Now, we need to consider the right chain $\Pi'$. The visibility
may be blocked by some vertex of $\Pi'$ at
the current position of $L$. So, we traverse the right chain from
$p_{i+1}$ to $p_j$. As soon as a blocking is observed, $L$ is modified as
mentioned in the algorithm. However, in the new position of $L$, a portion
of it becomes closer to the left chain $\Pi$; thus, it may again be obstructed
by $\Pi$. So, we  need to traverse $\Pi$. This
alternating process may continue until $p_j$ is reached along the right
chain. It needs to be mentioned that during this traversal, we have not
noticed whether $L$ is obstructed by some edge which is already visited.
So, finally we execute Pass-3 to check this. 

The time complexity follows from the fact that $\Pi$ is traversed three
times and $\Pi'$ is traversed once only. During the traversal of $\Pi$
(resp. $\Pi'$) its each vertex is visited at most once. \qed 
\end{proof}
\vspace{-0.2in}

\begin{theorem} \label{wv}
The weak visibility polygon of an edge of $P$
can be computed in $O(n^2)$ time using $O(1)$ extra space,
where the vertices of $P$ are given in a read-only
array. 
\end{theorem}

\vspace{-0.3in}

%%%%%%%%%%%%%%%%%%%%%%%%%%% MINIMUM LINK PATH  %%%%%%%%%%%%%%%%%%%%%%%%%%%%
%%%%%%%%%%%%%%%%%%%%%%%%%%%%%%%%%%%%%%%%%%%%%%%%%%%%%%%%%%%%%%%%%%%%%%%%%%%

\section{Minimum link path between a pair of points} \vspace{-0.1in}
Given a polygon $P$ and a pair of points $s$ and $t$, the {\it minimum link 
path} between  $s$ and $t$, denoted by $MLP(s,t)$, is a polygonal chain from 
$s$ to $t$ where the number of edges in the chain is minimum among all other 
polygonal chains connecting $s$ and $t$ (Figure \ref{fig:wvp}(b)). We 
propose an algorithm for computing $MLP(s,t)$ assuming that the 
vertices of $P$ are given in a readonly array in anticlockwise order. 

A classic way to compute $MLP(s,t)$ is as follows: (i) Compute the
visibility
polygon of $s$, called $Q$. If $t\in Q$, then $s$ and $t$ are straight-line
visible. Otherwise (ii) identify the edge $\chi$ of $Q$ such that the
sub-polygons of $P$ lying on one side of $\chi$  contains $s$ and that on the
other side contains $t$. Now (iii) compute the weak visibility polygon of $\chi$
in the sub-polygon containing $t$. If again it contains $t$, then the process
stops; otherwise  iterate steps (ii) and (iii). The time complexity is $O(kn^2)$
and it needs $O(1)$ extra space (see Theorem \ref{wv}), where $k$ is the number
of segments in  $MLP(s,t)$.

We now describe an algorithm for this problem that runs in $O(kn)$ time and
$O(1)$ work-space. It also executes $k$ iterations to report the $k$ segments of
the path. We use a variable $\pi$ to store the point such that the path from $s$
to $\pi$ is already reported. Initially $\pi$ stores $s$. In each iteration we
execute the procedures $\LNV(e,e')$ described in Section \ref{LNV}. It is
tailored such that it can work even if $e$ and/or $e'$ is/are point(s) inside
$P$, and returns two parameters $\Phi$ and $L$, where $\Phi$ is the left non-visible 
portion of $e$ from $e'=t$ and $L$ is a line such that nothing to the left of $L$ 
is visible to $e'=t$. If $e$ is a point, then we draw a horizontal line segment $[a,b]$
inside $P$ that passes through $e$ and the points $a,b$ lying on the boundary of
$P$. Similarly, the line segment $[c,d]$ is defined for $e'$ if it is a point.
Initially, $e$ and $e'$ are the points $s$ and $t$, respectively. Here $\Pi$ is
defined as an anticlockwise polychain from $a$ to $c$ and $\Pi'$ is a clockwise
polychain from $b$ to $d$. From next iteration onwards $e=[a,b]$ will be an edge
of the corresponding polygon, but $e'$ remains equal to $t$ in all the
iterations.  

At each step, if $\Phi \neq e$ and one of the end-points of $L$ is equal to $t$,
then $t$ is visible from some point of $e$. We choose the left-most point $q$ of
$\Phi$, and report the last two edges $[\pi,q]$ and $[q,t]$ of the \MLP. \\If
$\Phi=e$, then no part of $e$ is visible from $t$; We execute the procedure
{\textsf Compute-First-Link}($\Phi,L$), stated below to compute an edge $\Psi$
of the weak-visibility polygon of $e$ such that 
$s$ and $t$ are in two different sides of $\Psi$. 
\vspace{-0.1in}
\begin{observation} \label{ob}
(i) Every point of $\Psi$ is weakly visible from $e$ and no point in the proper 
interior of the sub-polygon of $P$ containing $t$ is visible from $e$.\\
(ii) Extension of $\Psi$ intersects $e$ (at a point, say $q$).
\end{observation}
\vspace{-0.1in}
We report the link $[\pi,q]$; reset $\pi=q$,
and execute the next iteration with $e=\Psi$. 
\vspace{-0.25in}
\subsection{\textsf Compute-First-Link}\vspace{-0.1in}
Here we need to handle the following two cases: (i) $L$ does not intersect 
the edge $e$ but contains the point $t$, and (ii) $L$ does not contain the 
point $t$. 
\begin{figure}[t]
\centering
\includegraphics[scale=0.30]{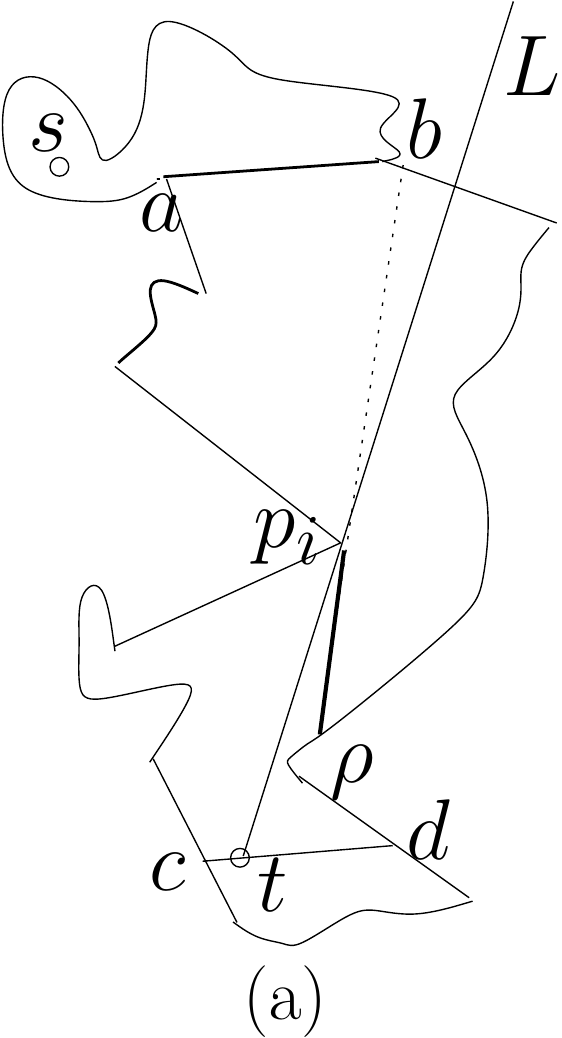}\hspace{0.2in}
\includegraphics[scale=0.30]{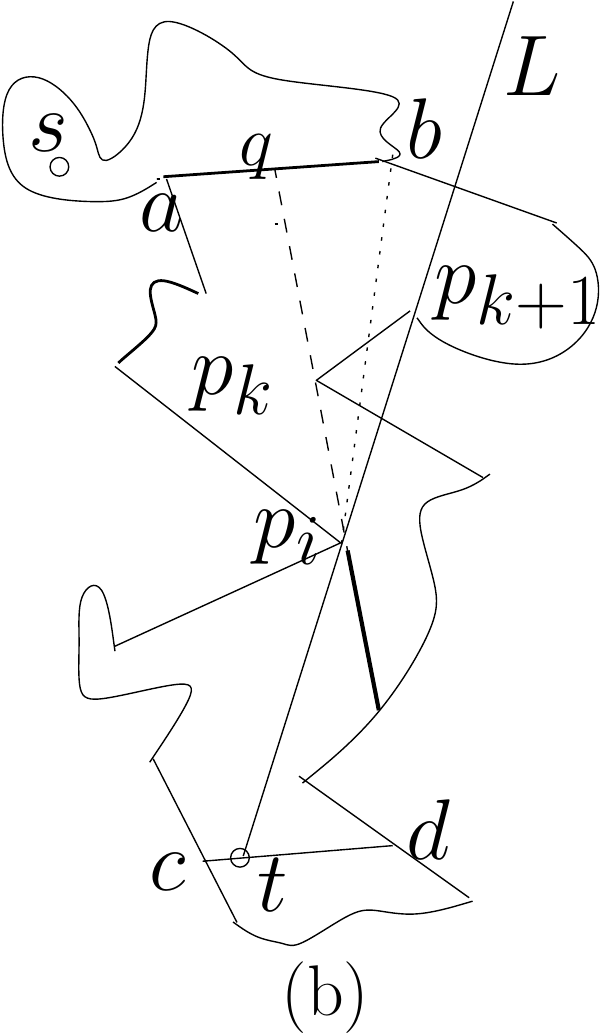}\hspace{0.2in}
\includegraphics[scale=0.30]{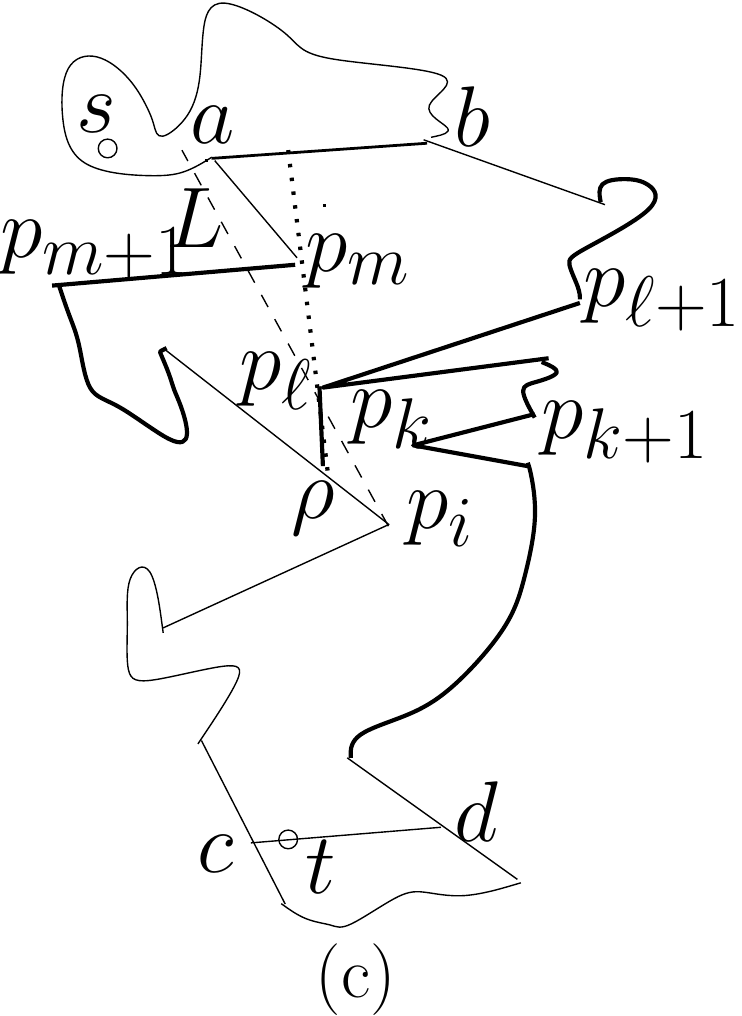}\hspace{0.2in}
\includegraphics[scale=0.30]{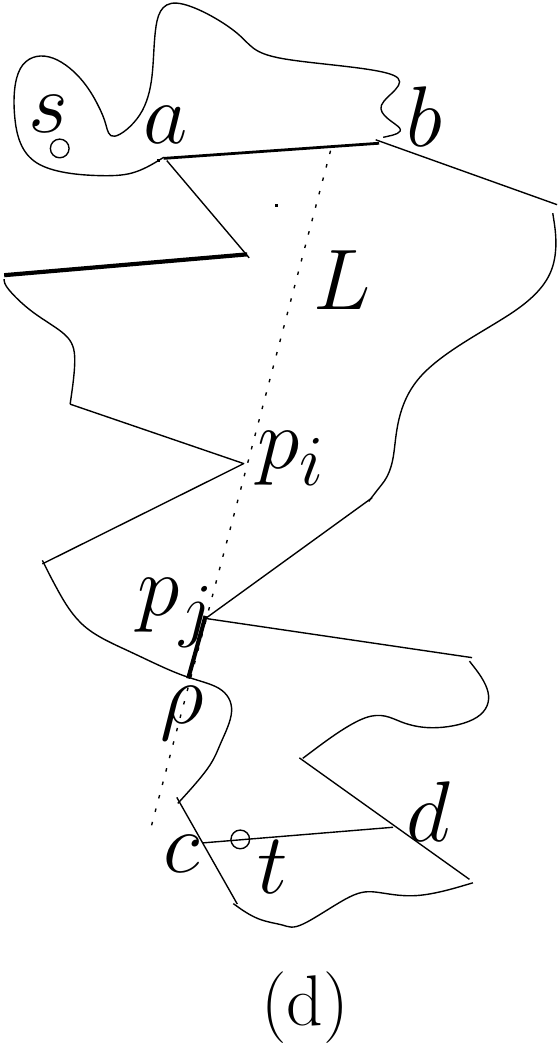}\hspace{0.2in}
\includegraphics[scale=0.30]{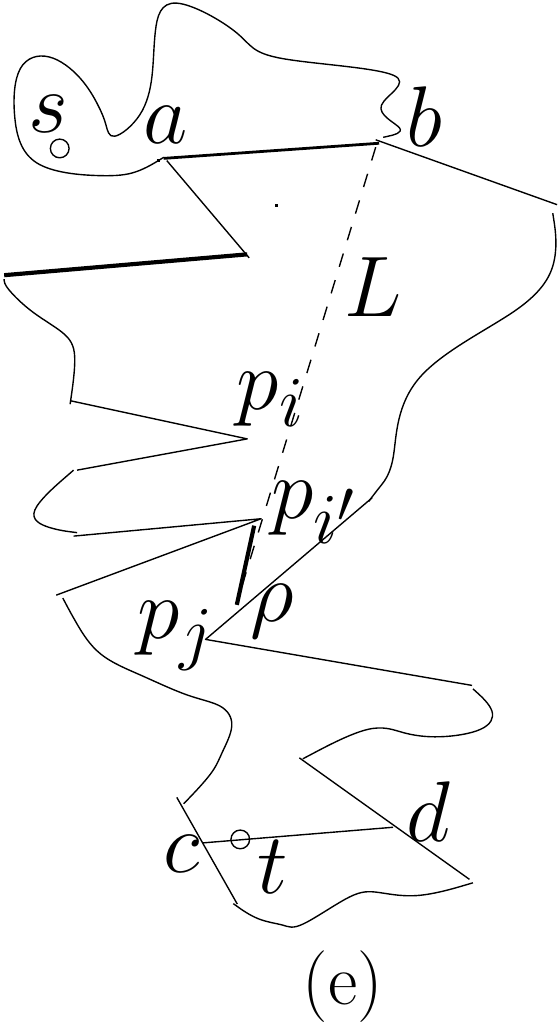}
\vspace{-0.1in}
\caption{Computation of minimum link path}\vspace{-0.3in}
\label{fig:mlpx}
\end{figure}

\vspace{-0.2in}
\subsubsection{Case (i)} 
Let $L$ be defined by a vertex $p_i \in \Pi$ and $t$ (see Figure
\ref{fig:mlpx}). We compute the right non-visible portion of $e=[a,b]$ from
the point $p_i$. We redefine $L=[b,p_i]$ and alternately visit the
vertices of the polychain $\Pi'$ from $d$ to $b$ in anticlockwise order and the
polychain $\Pi$ from $p_i$ to $a$ in clockwise order using the procedures {\bf
Process-1} and {\bf Process-2} stated below. Note that, $[c,d]$ is a chord of
$P$ containing $t$, as stated earlier. 

{\bf Process-1}:\vspace{-0.1in}
\begin{itemize}
\item[$\bullet$] If no edge of $\Pi'$ intersects $L$ above $p_i$ then it returns
$\Psi=[p_i,\rho]$, where $\rho$ is the point of intersection of $L$ and $\Pi'$
(see Figure \ref{fig:mlpx}(a)).  

\item[$\bullet$] If an edge $(p_k,p_{k+1})$ of $\Pi'$ intersects $L$, then $L$
is modified by $[p_i,p_k)$, and the end-point of $L$ on $e$ moves towards $a$.
For each encounter of an edge of $\Pi'$ intersecting $L$, $L$ is modified 
accordingly. The traversal continues along $\Pi'$ until $b$ is reached (Figure \ref{fig:mlpx}(b)) or 
an edge $(p_\ell,p_{\ell+1}) \in \Pi'$ is encountered such that the redefined 
$L$ does not intersect $e$ (Figure \ref{fig:mlpx}(c)). Now {\bf Process-2} 
is invoked. 
\end{itemize}
\vspace{-0.1in}
{\bf Process-2}:\\
We redefine $L=[p_\ell,a]$ and start traversing $\Pi$ from $p_i$ towards $a$ 
to compute the left non-visible portion of $e$ from the point $p_\ell$. As in
{\bf Process-1}, for each encounter of an edge $(p_m,p_{m+1}) \in \Pi$
intersecting $L$, the end-point of $L$ on $e$ moves towards $b$. Finally 
\vspace{-0.1in}
\begin{itemize}
\item[$\bullet$] If $a$ is reached and $L$ does not leave $e$, then report 
$\Psi=[p_{m+1},\rho]$, where $\rho$ is the point of intersection of $L$ and the
polychain $\Pi'$ (Figure \ref{fig:mlpx}(c)).

\item[$\bullet$] If the end-point of $L$ on $e$ goes beyond $b$ for an edge 
$(p_m,p_{m+1}) \in \Pi$, then $L$ is redefined as $[p_{m+1},b]$ and  
{\bf Process-1} is invoked.
\end{itemize}
\vspace{-0.2in}

\subsubsection{Case (ii)}
Let $L$ be defined by two vertices $p_i \in \Pi$ and $p_j \in \Pi'$. As in
the algorithm \LNV\, here also we traverse from $c$ to $p_i$ along $\Pi$ to get
a vertex that intersects $[p_j,p_i]$. If no such vertex is found and
$L$ intersects $\Pi$ at a point $\rho$, then $\Psi=[\rho,p_j]$ is returned (see
Figure \ref{fig:mlpx}(d)). 
If such a vertex $p_{i'}$ is found, then $L$ is redefined by
$p_j$ and $p_{i'}$, and the traversal from $p_j$ starts to get a vertex in
$\Pi'$ that intersects $[p_j,p_{i'}]$ (see Figure \ref{fig:mlpx}(e)). This type
alternate traversal in $\Pi$
and $\Pi'$ ultimately defines a line $L$ by two vertices $p_\alpha \in \Pi$ and
$p_\beta \in \Pi'$ such that there exists no edge $(p_k,p_{k+1})$ in $\Pi$
with $k > i$ that intersects $L$. Now if $L$ does not intersect $e$, the
situation is similar to Case (i). Otherwise, the traversal along
$\Pi'$ continues from $p_\beta$ until it is obstructed by a vertex
in  $\Pi$. Then the traversal starts from $p_\alpha \in \Pi$. This
type of alternate traversal in $\Pi$ and $\Pi'$ continues as in
Case (i) to have a chord of $P$ defined by its two vertices. 
  \vspace{-0.1in}
\begin{theorem}
The proposed algorithm correctly computes the {\em minimum link path} between
$s$ and $t$ in time $O(kn)$, where $P$ is given in a read-only array, and $k$ is
the size of the output. 
\end{theorem}\vspace{-0.1in}
\begin{proof}
The above scheme reports an edge $\eta$ of the visibility polygon of $e$ such
that $s$ and $t$ are in different sides of it. The minimality of the path length
follows from the fact that $\eta$ satisfies Observation \ref{ob}. The above
scheme needs $O(n)$ time since it needs visiting all the vertices of $\Pi$ and
$\Pi'$ at most a constant number of times. Since the length of the minimum link
path is $k$, the result follows.
\end{proof}

\vspace{-0.2in}
\small
\bibliographystyle{abbrv}
% or: plain,unsrt,alpha,abbrv,acm,apalike,.
\bibliography{references}

\newpage

\begin{center}
{\bf Appendix}
\end{center}

\begin{algorithm}[h!]
\scriptsize
\SetAlgoLined
\KwIn{A simple polygon in the array $P[]$  and a point $\pi$ inside the
polygon}
\KwOut{Visibility polygon of $\pi$}

Draw a horizontal line  $\overrightarrow{H}$ from $\pi$ towards right and
find the first intersection point $q$ with the polygon $P$. Let $q \in$
the edge $(p_{\theta},p_{\theta+1})$.\;

Initialize: $k=l= \theta+1$ ; $i=\theta+2$; $previous\_vertex=
P[\theta+1]$\;

\While{$i \neq \theta$}
{  (* Case 1: *)\;
  \If{$\angle q\pi P[\ell] <\angle q\pi P[i]$}
  { Push\_and\_Proceed($i,\ell$)\;}
   (* Case 2: *)\;
  \ElseIf{$\angle q\pi P[\ell] >\angle q\pi P[i]$ and the polygon makes a
   right turn at $P[i]$}
   {
   \While{$(\angle q\pi P[\ell] >\angle q\pi P[i]) \bigwedge (i\neq
   \theta)$} {Only\_Proceed($i$)}
   \If{$\angle q\pi P[\ell] <\angle q\pi P[i]$}
   {$\phi=$ intersection point of edge $(p_{i-1},p_i)$ and the
    half-line  $[\pi,P[\ell])$\;
Push\_Phi($\phi, \ell$)\;    PUSH\_and\_PROCEED($i,\ell$)\;
   }
  }
  (* Case 3: *)\;
  \Else
  {
    \While{$!(Case3.1 \vee Case3.2 \vee Case 3.3)$}
    {$\ell=\ell -1$\;}
    (* Case 3.1: *)\;
    \If{$\ell < k$} 
    {
      $k=i-1$; $P[k]=q$; $\ell=i$\; 
    Only\_Proceed($i$)
    }
    (* Case 3.2: *)\;
    \ElseIf{$\angle{q\pi P[\ell]} < \angle{q\pi P[i]}$}
    {
       $\phi=$ intersection point of edge $(P[\ell],P[\ell+1])$ and the
half-line $[\pi,P[i])$\;
Push\_Phi($\phi, \ell$)\;
        PUSH\_and\_PROCEED($i,\ell$)\; 	
    }
    (* Case 3.3: ($\angle q\pi P[\ell] > \angle q\pi P[i]$) $\wedge$
($[P[i-1],P[i]]$ and $[P[\ell],P[\ell+1]]$ properly intersect) *)\;
    \Else
    {
      \While{!$((\angle q\pi P[\ell] <\angle q\pi
P[i])\bigwedge(i=\theta))$} {Only\_Proceed($i$)}
      
      \If{$i\neq \theta$}
      {
      $\phi=$ intersection point of the edge $(previous\_vertex,P[i])$ and
the half-line $[\pi,P[\ell])$\;
      Push\_Phi($\phi, \ell$)\;
	PUSH\_and\_PROCEED($i,\ell$)	
      }
  }
}
}
Report $P[k \ldots \ell]$ of the array\;
end.\\
---------------------------------------------------\\
{\bf Procedure PUSH\_and\_PROCEED($i,\ell$)}\;
  $\ell=(\ell+1) mod\ n$;~~
      $previous\_vertex=P[i]$\;
      swap($P[i],P[\ell]$); ~~
       $i=(i+1) mod\ n$\;
end.\\
---------------------------------------------------\\
{\bf Procedure Only\_Proceed($i$)}\;
  $previous\_vertex=P[i]$; ~~$i=(i+1) mod\ n$\;
end.\\
---------------------------------------------------\\
{\bf Procedure Push\_Phi($\phi, \ell$)}\;
  $\ell=(\ell +1) mod\ n$; ~~	$P[\ell]=\phi$\;
end.
 \caption{\IV}
\label{Vis}
\normalsize
\end{algorithm}

\begin{algorithm}
\scriptsize
\SetAlgoLined
\SetKwInOut{Workspace}{WorkSpace}
\Workspace{An array $R$  of size   $2 \times \lceil \sqrt n \rceil $, and
array  $S$  of size  $\lfloor \sqrt n \rfloor$}

$R[1,j]$ and $R[2,j]$ contain the blocking vertices of $P_j$ from its two
sides\; 
While processing $P_j$, it is copied in $S$ and processed there.
The array $P$ remains unaltered\; 
Variables used\;
$\ell$: top pointer of stack maintained at the beginning of $S$\; 
$\chi$: index of the poly-chain of maximum index that is totally/partially
visible\;
$i$: index of the current vertex in $S$\;

(* Modified Case 3.1: Here $\ell$ is reached to 0 *)\;

\While{$(R[1,\chi]=0 \wedge R[2,\chi]=0) \vee
((previous\_vertex,S[i])$ intersects both the line $[\pi, 
p_\alpha)$ and $[\pi,p_\beta)) \vee (\chi \neq 0)$}
      {
	\If{$(previous\_vertex,S[i])$ intersects both the line $[\pi, 
            p_\alpha)$ and $[\pi,p_\beta)$}
          {  $(R[1,\chi]=0$;  $R[2,\chi]=0$\;  }
	$\chi=\chi-1$\;
     }
\If{$\chi \neq 0$}
{     \If{$(previous\_vertex,S[i])$ intersects none of the line $[\pi, 
          p_\alpha)$ and $[\pi,p_\beta)$}
    {

      $previous\_vertex= S[i]$; $\ell=\ell+1$; $S[\ell]=S[i]$; $i=i+1$\;

    }
    \Else{$R[2,\chi]=i$\;  $previous\_vertex= S[i]$; $\ell=\ell+1$;
$S[\ell]=S[i]$; $i=i+1$\;}

  }

\caption{Modified-Case-3.1}
\label{ARV}
\normalsize
\end{algorithm}

\begin{algorithm}
\scriptsize
\SetAlgoLined
$j=1$\;
\While{$j \neq No\_of\_Partition$}
{
  \If{$R[1,j] \neq 0 \wedge R[2,j] \neq 0$}
  {
    $\alpha=R[1,j]$; $\beta=R[2,j]$\;
     \If{$P[\alpha] \notin P_j$}
      {Find the edge $e$ of $P_j$ which is intersected first by the line
                   $[\pi,P[\alpha)$\;
        $\phi=$ intersection of $e$ and $[\pi,P[\alpha])$
      }
    
     \Else{$\phi=P[\alpha]$}

     \If{$P[\beta] \notin P_j$}
      {Find the edge $e$ of $P_j$ which is intersected first by the line
      $[\pi,P[\beta)$\;
      $\psi=$ intersection of $e$ and $[\pi,P[\beta])$}
     \Else{$\psi=P[\beta]$}

    Copy $\phi$ to $\psi$ of the polychain $S$\;
    Execute Algorithm \IV\ on $S$ to report the visible portion
  } 
$j=j+1$
}
\caption{Pass 2 of Algorithm \RV}
\normalsize
\end{algorithm}

\begin{algorithm}
\scriptsize
\SetAlgoLined
%\SetKwInOut{Workspace}{WorkSpace}
\KwIn{Two edges $p_ip_{i+1}$ and $p_jp_{j+1}$ of a  simple polygon $P$}
\KwOut{The left portion of $p_jp_{j+1}$ which is not weakly-visible from
$p_ip_{i+1}$}
%\Workspace{Array $R$  of size   $2 \times \lceil \sqrt n \rceil $, and}

$PASS1()$\;
$r=PASS2()$\;

\If{$r=-1$}
{
  Report nothing is visible\;
}
\Else
{$PASS3()$\;}
end.\\
---------------------------------------------------------\\
{\bf Procedure $PASS1()$}\; 

$L=(p_{j+1},p_{i+1})$\;
$t=(j+2) mod\ n$\;
\While{$t \neq i$}
{

  \If{$p_t$ is to the right of $L$}
  {
     {\sc UpdateLine}($p_t,p_{i+1}$)\;
     $\tau=t$ \;
  }
$t=(t+1) mod\ n$\;
}
end.\\
---------------------------------------------------------\\
{\bf Procedure $PASS2()$}\; 
$k=i+1$; $\ell=i$; $m=\tau$\;
$r=1$\;
\While{$r\neq0 \wedge r\neq-1$}
{
  \If{$r=1$}
  {$Process1()$\;}
  \ElseIf{$r=2$}{$Process2()$\;}

}
Return $r$\;
end.\\
---------------------------------------------------------\\
{\bf Procedure $Process1()$}

  \If{$k=j$}
  {$r=0$\;}
  \Else
  {
      \If{$(p_{k-1}p_k)$ does not intersect line $L$}
      {
	$k=(k+1) mod\ n $\;
	$r=1$\;
      }
      \ElseIf{$(p_{k-1}p_k)$  intersects below $p_m$ }
      {
      {\sc UpdateLine}($p_m,p_k$)\;
      $r=2$\;
      }
      \Else
      { $r=-1$\;}
  }
end.\\
---------------------------------------------------------\\
{\bf Procedure $Process2()$}

 \If{$l > \tau$}
  {

    \If{$p_{\ell+1}p_{\ell}$ does not intersect $L$}
    {$\ell=(\ell-1) \bmod{n}$\; $r=2$\;}
    \ElseIf{$p_{\ell+1}p_{\ell}$  intersects $L$ below $p_k$}
    {$r=-1$\;}
    \Else
    {
         {\sc UpdateLine}($p_l,p_k$)\;
	 $m=\ell$ \;
	 $r=1$ \;
    }
  }
\Else{$r=0$\;}
end.\\
---------------------------------------------------------\\
{\bf Procedure $PASS3()$} 

$t=(j+2) mod\ n$\;
\While{$(t \neq i) \bigwedge (r \neq -1)$}
{

  \If{$p_t$ is to the right of $L$}
  {
    $r=-1$\;
  }
$t=(t+1) mod\ n$\;
}

$t=(i+2) mod\ n$\;

\While{$(t \neq j) \bigwedge (r \neq -1)$}
{

  \If{$p_t$ is to the left of $L$}
  {
       $r=-1$\;
  }
$t=(t+1) mod\ n$\;
}

\If{$r=-1$}
{Report nothing is visible\;}
\Else{Report $(p_{j+1},\theta')$ is not visible\;}
end.
\caption{$\LNV(v_iv_{i+1},v_jv_{j+1})$}
\label{LNV}
\normalsize
\end{algorithm}

\end{document}